\documentclass[aps,prd,showpacs,nofootinbib,preprintnumbers,twocolumn]{revtex4-1}
\usepackage{amsmath,amssymb}
\usepackage[dvips]{graphicx}

\usepackage[bookmarks=true,bookmarksnumbered=true,
colorlinks=false,pdftitle={Paper},pdfauthor={Mananori Hanada,
Yoshinori Matsuo, Naoki Yamamoto},
pdfsubject={},pdfkeywords={}]{hyperref}
\usepackage{enumerate}
\newcommand \beq{\begin{eqnarray}}
\newcommand \eeq{\end{eqnarray}}

\newcommand{\SU}{\text{SU}}
\newcommand{\SO}{\text{SO}}
\newcommand{\Sp}{\text{Sp}}
\newcommand{\U}{\text{U}}

\DeclareMathOperator{\tr}{Tr}
\def\simge{\mathrel{
       \rlap{\raise 0.511ex \hbox{$>$}}{\lower 0.511ex \hbox{$\sim$}}}}
\def\simle{\mathrel{
       \rlap{\raise 0.511ex \hbox{$<$}}{\lower 0.511ex \hbox{$\sim$}}}}

\begin{document}
\preprint{KEK-CP-273, KEK-TH-1541, NSF-KITP-12-065, INT-PUB-12-020, YITP-12-35}
\title{Sign problem and phase quenching in finite-density QCD:\\
models, holography, and lattice}

\author{Masanori Hanada$^{1,2}$}
\author{Yoshinori Matsuo$^1$}
\author{Naoki Yamamoto$^{3,4}$}
\affiliation{
$^1$KEK Theory Center, 
High Energy Accelerator Research Organization (KEK), 
Tsukuba 305-0801, Japan\\
$^2$Kavli Institute for Theoretical Physics, University of California,
Santa Barbara, CA 93106-4030, USA\\
$^3$Institute for Nuclear Theory, 
University of Washington, Seattle, WA 98195-1550, USA\\
$^4$Yukawa Institute for Theoretical Physics,
Kyoto University, Kyoto 606-8502, Japan}
\date{\today}

\begin{abstract}
The effect of the complex phase of the fermion determinant is a key question 
related to the sign problem in finite-density QCD. 
Recently it has been shown that 
ignoring the complex phase -- the phase quenching -- 
does not change physics in a certain region of 
the phase diagram when a number of colors $N_c$ is large. 
In this paper we study the effect of the phase quenching 
within the frameworks of effective models and holographic models.
We show, in a unified manner, that the phase quenching gives exact results for any
fermionic observables (e.g., chiral condensate) in the mean-field approximation 
and for gauge-invariant gluonic observables (e.g., Polyakov loop)
to one-meson-loop corrections beyond mean field.
We also discuss implications for the lattice simulations and confirm 
good quantitative agreement between our prediction and existing lattice QCD results. 
Therefore the phase quenching provides rather accurate answer already at $N_c=3$ 
with small $1/N_c$ corrections which can be taken into account by the phase reweighting.

\end{abstract}
\pacs{12.38.Gc, 11.15.Pg, 12.38.Lg, 12.38.Mh}
\maketitle

\tableofcontents

\section{Introduction and summary}
\label{sec:introduction}
Phases of matter under extreme conditions, such as the hottest matter in 
the early Universe and relativistic heavy ion collisions, 
and the most dense matter inside the core of neutron stars, 
are described by Quantum Chromodynamics (QCD) 
at finite temperature and/or finite density.
Due to the strong-coupling nature of QCD, lattice simulations 
based on importance sampling have been the main first 
principle method to reveal the properties of such systems.
A number of important properties of hot QCD matter, such as the equation of state 
\cite{Borsanyi:2010cj} and a rapid crossover from hadronic matter to quark matter
\cite{Aoki:2006we}, have been unraveled near zero chemical potential.  
However, studies of dense QCD matter are difficult because of the notorious  {\it sign problem}.

Recently it has been realized that 
QCD at finite baryon chemical potential $\mu_B$ (QCD$_B$) is equivalent to 
QCD at finite isospin chemical potential $\mu_I$ (QCD$_I$),\footnote{In 
this paper we consider the two-flavor QCD unless otherwise stated. 
The baryon chemical potential $\mu_B$ means $\mu_1=\mu_2=\mu=\mu_B/N_c$, 
while the isospin chemical potential $\mu_I$ stands for $\mu_1=-\mu_2=\mu=\mu_I/2$. 
We also assume the degenerate quark mass so that the isospin symmetry is exact.}
with a large number of colors $N_c$ in a certain region of the 
phase diagram presumably relevant to the heavy ion collision experiments
\cite{Cherman:2010jj,Hanada:2011ju} (see also below). 
Because QCD$_I$ does not suffer from the sign problem \cite{Alford:1998sd, Son:2000xc}, 
this equivalence enables us to study properties of QCD$_B$
through the lattice simulation of QCD$_I$.

\begin{figure}[t]
\begin{center}
\includegraphics[width=7cm]{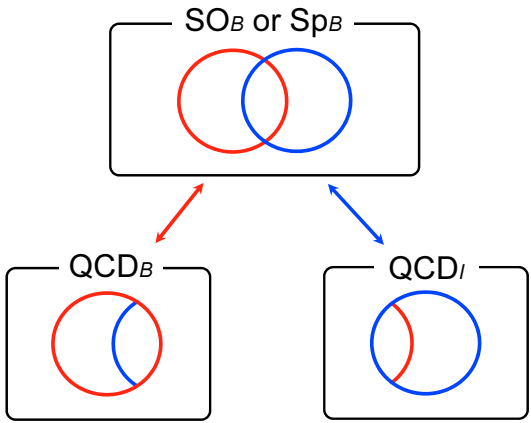}
\end{center}
\vspace{-0.5cm}
\caption{
Relations between $\SO(2N_c)$ [or $\Sp(2N_c)$] gauge theory 
at finite $\mu_B$ ($\SO_B$ or $\Sp_B$), 
QCD at finite $\mu_B$ (QCD$_B$), 
and QCD at finite $\mu_I$ (QCD$_I$). 
A class of correlation functions in $\SO(2N_c)$ theory 
[$\U(1)_B$-neutral operators; red circle] coincide with the counterparts in QCD$_B$ and 
another class of those (isospin singlets; blue circle) coincide with the counterparts in QCD$_I$.
There is an intersection of these two classes in $\SO(2N_c)$, which leads to 
the equivalence between the counterparts in QCD$_B$ and QCD$_I$.
For the detail, see Ref.~\cite{Hanada:2011ju}.}
\label{fig:equivalence}
\end{figure}

This equivalence has been derived by using a string-inspired large-$N_c$ technique   
known as the {\it orbifold equivalence} \cite{Kachru:1998ys,Lawrence:1998ja,
Bershadsky:1998mb,Bershadsky:1998cb,Kovtun:2004bz}.
(For the idea of the orbifold equivalence, see Sec.~\ref{sec:orbifolding}.)
As shown in Refs.~\cite{Cherman:2010jj,Hanada:2011ju}, 
there are ${\mathbb Z}_2$ projections called the {\it orbifold projections} relating 
$\SO(2N_c)$ and $\Sp(2N_c)$ gauge theories\footnote{The symplectic group is defined as 
$\Sp(2N_c) = \{g \in \U(2N_c) |g^T J_c g = J_c \}$, where $J_c = -i\sigma_{2} \otimes 1_{N_{c}}$.
This is also denoted as USp$(2N_c)$.} at finite $\mu_B$ 
($\SO_B$ and $\Sp_B$) to QCD$_B$ and QCD$_I$.
The relations between these theories are summarized in Fig.~\ref{fig:equivalence}.
The large-$N_c$ orbifold equivalence guarantees that these theories 
are equivalent in the sense that a class of correlation functions
(e.g., magnitude of the chiral condensate)
and the phase diagrams characterized by such quantities coincide, 
as long as the projection symmetry is not broken spontaneously \cite{Kovtun:2004bz}.  
This requirement is always satisfied for the equivalences between 
$\SO_B$, $\Sp_B$, and QCD$_I$. 
Although the projection symmetries relating these three theories to QCD$_B$ are broken 
spontaneously in the pion or diquark condensation phase 
(see Sec.~\ref{sec:phase} below), 
the equivalence holds outside that region.

The purposes of the present paper are two-fold.
Firstly, we develop a technique of the orbifold equivalence within 
the frameworks of effective models and holographic models
which are widely used to study the properties of finite-density QCD. 
The effective models covered in this paper include
the Nambu--Jona-Lasinio (NJL) model \cite{Nambu:1961tp},
linear sigma model (L$\sigma$M) \cite{GellMann:1960np},
Polyakov-Nambu--Jona-Lasinio (PNJL) model \cite{Fukushima:2003fw, Megias:2004hj, Ratti:2005jh}, 
Polyakov-quark-meson (PQM) model \cite{Schaefer:2007pw},
chiral random matrix model ($\chi$RMM) \cite{Shuryak:1992pi},
and strong-coupling expansion of lattice QCD \cite{Kogut:1974ag}.
The holographic models include the D3/D7 model \cite{Karch:2002sh} 
and Sakai-Sugimoto model \cite{hep-th/0412141}. 
We also explain implications of the orbifold projections 
for lattice QCD methods, such as the reweighting method \cite{Ferrenberg:1988yz, Barbour:1997ej}, 
QCD with imaginary chemical potential \cite{Alford:1998sd,de Forcrand:2002ci}, 
and Taylor expansion method \cite{Allton:2002zi, Allton:2005gk, Gavai:2008zr}.
Then we point out that, in previous simulation results, 
the phase quenching is found to be well satisfied
at $N_c=3$ for small-$\mu$ and high-$T$ region.
Secondly, as a consequence of the orbifold equivalence, we 
provide criteria for the validity of the phase quenching 
in a unified manner, independently of the effective models and lattice methods.
When these criteria are satisfied, there is no overlap problem 
(see Sec.~\ref{sec:overlap} for details).
Our criteria are summarized as follows:\footnote{We note here that, 
in the chiral random matrix model \cite{Klein:2003fy} and NJL model \cite{Toublan:2003tt}, 
the exactness of the phase quenching was already observed 
by computing the effective potentials in terms of the chiral condensate. 
In this paper, we rather provide the underlying principle why this should be so, 
regardless of the details of models. As a by-product, we can also predict the exactness 
of the phase quenching in other models (L$\sigma$M etc.) which has not yet 
been pointed out, to our best knowledge.}
\begin{enumerate}
\item
{{\it For which quantities?}---For any physical (gauge-invariant) observable. }

\item
{{\it To what extent?}---For fermionic quantities, the phase quenching 
is exact at the leading order in $1/N_c$ (planar and one fermion loop) 
in the large-$N_c$ QCD, and in the mean-field approximation (MFA)
in the effective models. For gluonic quantities, the phase quenching
is exact to the next-to-leading order in $1/N_c$ (planar and one fermion loop)
in the large-$N_c$ QCD, and to the one-meson-loop corrections beyond the MFA 
in the effective models.}

\item
{{\it In what region of the phase diagram?}---The phase quenching 
is valid outside the pion condensation phase of the corresponding
phase-quenched theory.}
\end{enumerate}

Before discussing a general but rather mathematical derivation of the above statement
based on the orbifold equivalence, let us explain a heuristic derivation for a specific observable 
at perturbative level \cite{Cohen:2004mw, Toublan:2005rq}.
As an example, let us consider the chiral condensate $\langle \bar{\psi} \psi \rangle
=\langle \bar{u} u \rangle + \langle \bar{d} d \rangle$.  
Each flavor is assumed to have the quark chemical potential $\mu_u$ and $\mu_d$.
At large $N_c$, the chiral condensate is dominated by one-fermion-loop planar diagrams,  
with no additional fermion loops attached. 
If the flavors are not mixed, the 
contribution to the chiral condensate is given by the summation 
of each flavor,
$\langle \bar \psi \psi \rangle_{\mu_u, \mu_d} = f(\mu_u) + f (\mu_d)$ with some
function $f(\mu)$. Here note that, as long as the ground state does not mix the flavors, 
the flavor mixing arises only through the diagram with additional fermion loop(s)
which is suppressed at large $N_c$.
Also note that $f(\mu)$ is an even function of $\mu$, $f(\mu)=f(-\mu)$, 
due to the charge conjugation symmetry.
Then the chiral condensate at finite $\mu_B$,
$\langle \bar \psi \psi \rangle_{\mu_u=\mu_d=\mu}$,
turns out to be equal to the chiral condensate at finite $\mu_I$,
$\langle \bar \psi \psi \rangle_{\mu_u=-\mu_d=\mu}$
at the leading order of $1/N_c$, 
\beq
\langle \bar \psi \psi \rangle_{\mu_B}
= f(\mu) + f(\mu) = f(\mu) + f(-\mu)
= \langle \bar \psi \psi \rangle_{\mu_I},
\eeq
and hence the phase quenching is exact in the large-$N_c$ limit.
As is found from this argument, the essence of the phase quenching
is the flavor decoupling of chemical potentials.\footnote{Precisely speaking, 
the flavor decoupling is a {\it sufficient} condition for the exact phase quenching. 
Even if there is a flavor mixing as $\langle \bar \psi \psi \rangle_{\mu_u, \mu_d} 
= f(\mu_u) + f (\mu_d) + g(\mu_u, \mu_d)$ with some function $g$, 
the phase quenching can be exact as long as $g(\mu_u, \mu_d)=g(\mu_u, -\mu_d)$ 
is satisfied.}
However, flavor decoupling is not satisfied if there is some
mixing between up and down quarks in the ground state.
This actually happens in the pion condensation phase of QCD$_I$, 
as we shall see in Sec.~\ref{sec:phase}.

It should be remarked that 
the arguments based on the orbifold equivalence are more general. 
They lead to criteria for the validity of the phase quenching systematically
(criteria for the validity on which correlation functions 
and which regions of the phase diagram) 
based on the projections and the symmetry breaking patterns in a unified manner. 
Furthermore, although the proof given in Ref.~\cite{Hanada:2011ju} 
applies to all orders in perturbation, there are convincing arguments 
that the orbifold equivalence holds nonperturbatively, 
based on the weak-coupling calculation at high density \cite{Hanada:2011ju},  
effective theory analysis \cite{Cherman:2011mh}, 
and holographic analogue \cite{Hanada:2012nj}.  

This paper is organized as follows. 
We start with reviewing the sign problem and the phase quenching in QCD$_B$
in Sec.~\ref{sec:phase_quench_finite_Nc}. 
After explaining the sign problem and the phase quenching 
in Sec.~\ref{sec:phase-quench}, 
we show the phase diagram of the phase-quenched QCD in Sec.~\ref{sec:phase}. 
In Sec.~\ref{sec:large-Nc} we argue the large-$N_c$ equivalence which 
assures the exactness of the phase quenching for various observables. 
In the following sections we consider implications of this equivalence to 
effective models of QCD (Sec.~\ref{sec:model}), 
holographic models (Sec.~\ref{sec:holography}), and 
lattice QCD (Sec.~\ref{sec:lattice}). 
Section~\ref{sec:discussion} is devoted to discussions and outlooks.  

\section{Phase-quenched QCD}\label{sec:phase_quench_finite_Nc}
In this section, we recapitulate the notion of the phase quenching.  
We also review the phase diagrams of the phase-quenched QCD (QCD$_I$),
$\SO_B$ and $\Sp_B$ which are important for later discussions
on the applicability of the phase quenching.
\subsection{Sign problem and phase quenching}
\label{sec:phase-quench}
We consider mass-degenerate two-flavor QCD at 
a finite baryon chemical potential $\mu_B=N_c \mu$.
The Lagrangian in the Euclidean spacetime is given by
\beq
\label{eq:Lagrangian}
{\cal L}_{\rm QCD} &=& \sum_{f=1}^2 \bar{\psi}_f
D(\mu) \psi_f + {\cal L}_{\rm YM},
\nonumber \\
{\cal L}_{\rm YM} &=& \frac{1}{4 g^{2} } \tr ({F}_{\mu \nu})^2,
\eeq
where $D(\mu)=\gamma^{\mu}D^{\mu} + m + \mu \gamma^4$
and $D^{\mu} = \partial^{\mu} + ig A^{\mu}$.
$\psi_f$ is the quark field with mass $m$ 
in the fundamental representation and
$A^{\mu}=A_a^{\mu} T^a$ is the gauge field with
$T^a$ being the $\SU(N_c)$ color generators.
The partition function reads
\beq
Z_{B}=\int dA \ [\det{D(\mu)}]^2 e^{-S_{\rm YM}},
\eeq
where $S_{\rm YM}$ is the action of the pure Yang-Mills.
The path integral measure of the theory is positive semi-definite only 
at $\mu_B=0$. This can be understood as follows:
if we define the eigenvalue of the operator 
$\gamma^{\mu}D^{\mu} + \mu \gamma^4$ as $i \lambda_n$,
it also has the eigenvalue $-i \lambda_n$ for $\lambda_n \neq 0$
due to the chiral symmetry (i.e. this operator anticommutes with $\gamma_5$).
Because $\gamma^{\mu}D^{\mu}$ is anti-Hermitian, 
$\lambda_n$ are real when $\mu=0$. 
This is no longer true 
at $\mu \neq 0$ where $\lambda_n$ are complex in general, $\lambda_n \in {\mathbb C}$.
Recalling that eigenvalues of $D(\mu)$ appear in pairs 
$(\pm i \lambda_n + m)$ for $\lambda_n \neq 0$, 
\beq
\label{eq:positivity}
\det{D(\mu)} = 
\prod_n (i \lambda_n + m)(-i \lambda_n + m)
=\prod_n (\lambda_n^2 + m^2), \nonumber \\
\eeq
is positive semi-definite (complex) at $\mu=0$ ($\mu \neq 0$). 
The complex fermion determinant at $\mu \neq 0$ is 
the notorious fermion sign problem, which prevents us from applying 
the Monte Carlo methods.  

One can think of the phase-quenched QCD where the complex phase of 
the fermion determinant in QCD$_B$ is ignored. 
(This is different from the usual {\it quenched} approximation in the sense that 
the absolute value of the fermion determinant is taken into account.) 
This theory does not have the sign problem by definition, and thus, 
it can be analyzed by Monte Carlo simulations, 
although its relation to QCD$_B$ is not clear {\it a priori}.
The partition function of the phase-quenched QCD is given by\footnote{The 
partition functions $Z_B$ and $Z_I$ are also denoted as $Z_{1+1}$ and 
$Z_{1+1^*}$ in the literature. Here $1^*$ stands for the so-called 
conjugate quark which has the chemical potential $-\mu$ 
(while the usual quark has the chemical potential $+\mu$) \cite{Stephanov:1996ki}.}
\beq
Z_{I}=\int dA \ |\det{D(\mu)}|^2 e^{-S_{\rm YM}}.
\eeq
The reason why we use $I$ in the subscript is that physically
this theory corresponds to QCD at finite {\it isospin} 
chemical potential $\mu_I=2\mu$  (QCD$_I$) \cite{Son:2000xc}. 
This can be understood by recalling that the fermion 
determinant of QCD$_I$ is given by
\beq
\det{D(\mu)} \cdot \det{D(-\mu)} = |\det{D(\mu)}|^2. 
\eeq 
Here the equality above follows from  
$\det{D(-\mu)} = [\det{D(\mu)}]^*$, which can be checked by using
\beq
\gamma_5 (\gamma^{\mu}D^{\mu} + m - \mu \gamma^0) \gamma^5
= (\gamma^{\mu}D^{\mu} + m + \mu \gamma^0)^{\dag}.
\eeq

The expectation value of an observable ${\cal O}$ in each theory is given by
\beq
\langle {\cal O} \rangle_{B} = \frac{1}{Z_B}\int dA \ {\cal O} \left(\det{D(\mu)}\right)^2 e^{-S_{\rm YM}}, \nonumber \\
\langle {\cal O} \rangle_{I} = \frac{1}{Z_I}\int dA \ {\cal O} |\det{D(\mu)}|^2 e^{-S_{\rm YM}}.
\eeq
Although one cannot calculate $\langle {\cal O}\rangle_B$ directly because of the sign problem, 
in principle one can calculate it by using a trivial relation 
\begin{eqnarray}
\langle {\cal O}\rangle_B
=
\frac{\langle {\cal O}e^{2i\theta}\rangle_I}{\langle e^{2i\theta}\rangle_I}, 
\end{eqnarray}
where $e^{i\theta}\equiv\det D(\mu)/|\det D(\mu)|$ is the phase factor of the fermion determinant. 
This approach is called the {\it phase reweighting}. 
In practice, however, both the numerator $\langle {\cal O}e^{2i\theta}\rangle_I$ and the denominator $\langle e^{2i\theta}\rangle_I$ 
becomes almost zero and it is impossible to study QCD$_B$ by using the reweighting method with a reasonable computational cost. 

\begin{figure}[t]
\begin{center}
\includegraphics[width=5cm]{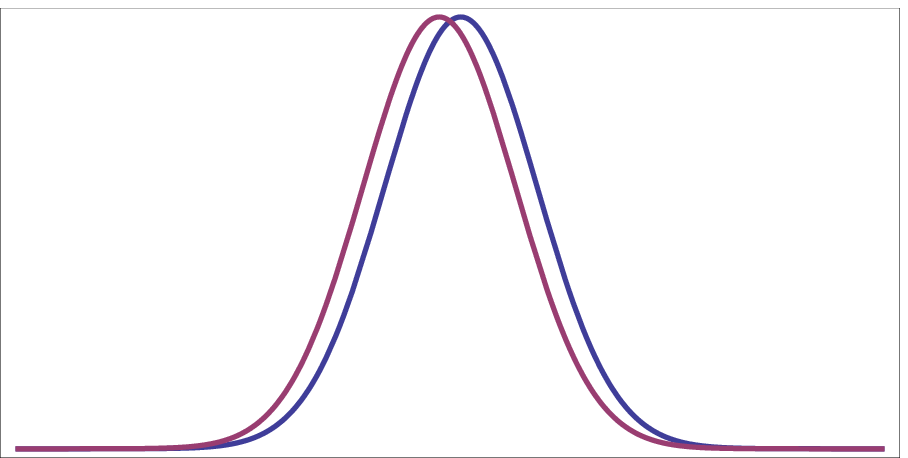}
\includegraphics[width=5cm]{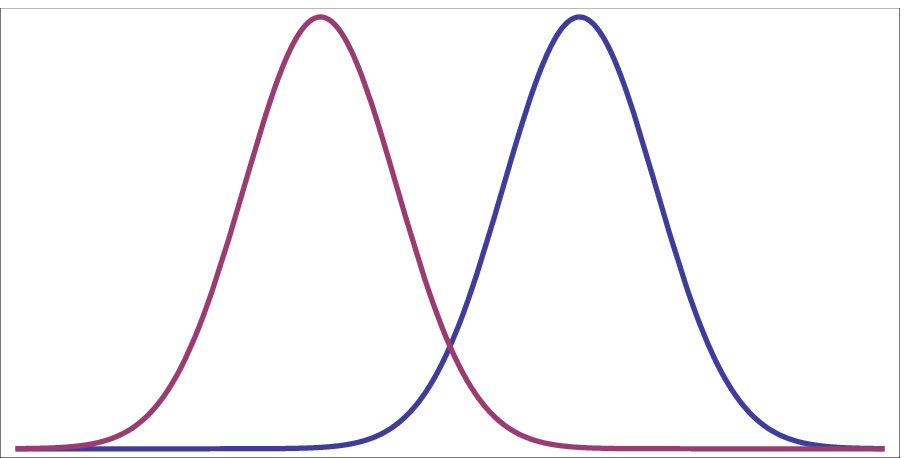}
\caption{Cartoons for the overlap problem. 
The horizontal axis stands for the value of an observable (e.g., the chiral condensate) 
and the blue and purple lines represent the path-integral weights of the full and 
phase-quenched theories. 
In the top figure, two distributions have large overlap. 
In finite-density QCD, it corresponds to a class of operators 
satisfying the criteria 1 in Sec.~\ref{sec:introduction}
in the small-$\mu$ region 
(where the pion does not condense in the phase-quenched theory). 
In the bottom figure, two configurations almost do not overlap. 
This corresponds to the large-$\mu$ region
and/or observables which do not satisfy the criteria 1.} 
\label{fig:overlap}
\end{center}
\end{figure}

Another related issue is that the phase-quenched ensemble (QCD$_I$) 
may not have large enough overlap with the ensemble in the full theory (QCD$_B$)
so that the importance sampling fails; for example, it might be possible 
that the peak in the phase-quenched ensemble disappears because of the phase fluctuation 
and the tail might correspond to the peak of the full theory (Fig.~\ref{fig:overlap}, right). 
If this is the case, in practice the configurations around the true vacuum do not appear at all. 
This problem is called the {\it overlap problem}. 
In QCD$_I$, as one increases the chemical potential, 
the pion condensation appears at some point (see Sec.~\ref{sec:phase}). 
There QCD$_I$ has completely different vacuum structure from QCD$_B$, 
and the severe overlap problem appears.
(Even outside the pion condensation, the overlap problem is absent only
for a class of observables. We will see that the orbifold equivalence
provides us with the criteria given in Sec.~\ref{sec:introduction}.)

The average phase factor $\langle e^{2i\theta}\rangle_I$ serves as a 
measure of the severity of the sign problem \cite{Splittorff:2006fu, Han:2008xj}; 
If the sign problem is mild (severe), it is close to unity (zero). 
Note however that  $\langle e^{2 i \theta} \rangle_{I} \sim 0$ does not necessarily exclude  
$\langle {\cal O} \rangle_{B} = \langle {\cal O} \rangle_{I}$.  
It happens when the phase and the observable factorize, 
$\langle {\cal O}e^{2i\theta}\rangle_I
=
\langle {\cal O}\rangle_I\cdot \langle e^{2i\theta}\rangle_I$.
If it is realized, we can compute some
$\langle {\cal O} \rangle_{B}$  
by computing $\langle {\cal O} \rangle_{I}$ without suffering from the sign problem,  
despite a vanishingly small average phase factor. 
As we shall show below, this actually happens 
in finite-density QCD in the large-$N_c$ limit and effective
models in the mean-field approximation.

\subsection{Phase diagrams of QCD$_I$, SO$_B$, and Sp$_B$}
\label{sec:phase}
\begin{figure}[t]
\begin{center}
\includegraphics[width=8cm]{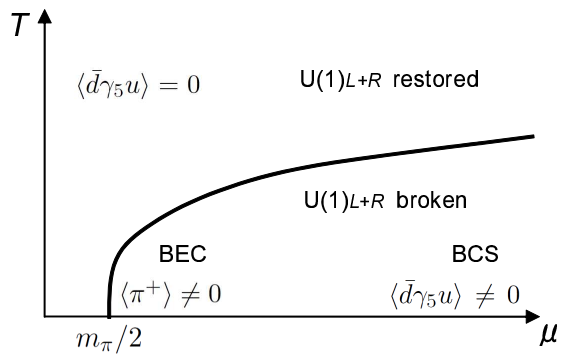}
\end{center}
\vspace{-0.5cm}
\caption{Phase diagram of phase-quenched QCD 
at finite $\mu$ (or QCD at finite $\mu_I = 2\mu$). 
Figure taken from Ref.~\cite{Hanada:2011ju}.}
\label{fig:muI}
\end{figure}

\begin{figure}[t]
\begin{center}
\includegraphics[width=8cm]{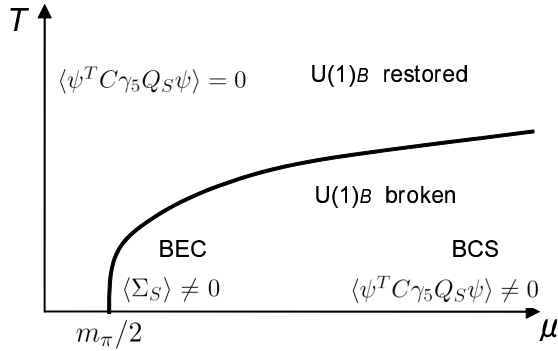}
\end{center}
\vspace{-0.5cm}
\caption{Phase diagram of $\SO(2N_c)$ gauge theory 
at finite $\mu$. 
Figure taken from Ref.~\cite{Hanada:2011ju}.}
\label{fig:SO}
\end{figure}

In this subsection we discuss the phase diagrams of the phase-quenched QCD
(QCD$_I$),  SO$_B$ and Sp$_B$,  with particular emphasis on the symmetry breaking patterns.
The phase diagram of QCD$_I$ in the $T$-$\mu_I$ plane was first investigated 
in Ref.~\cite{Son:2000xc}. It was shown recently \cite{Hanada:2011ju} that,
in the large-$N_c$ limit, $\SO_B$ and $\Sp_B$ have exactly 
the same phase structures as QCD$_I$.\footnote{In the large-$N_c$ limit and the chiral limit,
the phase diagram of QCD$_I$ also coincides with that of QCD at a finite
chiral chemical potential $\mu_5$ \cite{Hanada:2011jb}, 
where $\mu_5$ is defined as the chemical potential 
associated with the $\U(1)$ axial charge.}
Even at finite $N_c$, they are qualitatively the same. 
The phase diagrams of QCD$_I$ and $\SO_B$ are depicted in Fig.~\ref{fig:muI}
and Fig.~\ref{fig:SO}, respectively.

We first consider the ground state of QCD$_I$ at $T=0$.
As $\mu_I$ is increased, what happens first is the excitation of the 
lightest particle with the isospin number, the charged pion. 
At  $\mu_I>m_{\pi}$ ($\mu>m_{\pi}/2$), where $m_{\pi}$ is the pion mass in the QCD vacuum, 
the excitation energy of a charged pion, $m_{\pi}-\mu_I$, 
becomes negative and it is energetically favorable to excite it.  
Since a pion is a boson, this is the Bose-Einstein condensation 
(BEC) of pions which breaks $\U(1)_{L+R}$ down to ${\mathbb Z}_2$.

In the opposite limit, at sufficiently large $\mu_I$, 
interactions around the Fermi surface become weak because of the asymptotic freedom,  
and fundamental degrees of freedom are quarks and gluons.
The one-gluon exchange interaction between quarks near the 
Fermi surface leads to an attractive interaction in the 
color singlet channel. According to the Bardeen-Cooper-Schrieffer (BCS) mechanism, 
the Cooper pairing 
$\langle \bar d \gamma_5 u \rangle$ is formed \cite{Son:2000xc}.\footnote{Precisely
speaking, the one-gluon exchange interaction does not distinguish
between the condensates $\langle \bar d \gamma_5 u \rangle$ and 
$\langle \bar d u \rangle$. The condensate $\langle \bar d \gamma_5 u \rangle$
is favored by the nonperturbative instanton-induced interaction \cite{Son:2000xc}.}
Because the quantum numbers of condensates and 
symmetry breaking patterns are same in the BEC and BCS regimes, 
it is plausible that these two regimes are smoothly connected without any phase transition. 
This is the so-called BEC-BCS crossover. 

As $T$ is increased, the pion condensation is melted away and 
$\U(1)_{L+R}$ symmetry is restored.
In the BEC regime, the critical chemical potential $\mu_c$
for the pion condensation is not so sensitive to $T$ at low temperature. 
In the BCS regime, the critical temperature $T_c$ should be an 
increasing function of $\mu$, since the phase space near the 
Fermi surface for the pairing becomes larger with increasing $\mu_I$.
A natural scenario to continuously connect these two regimes is 
shown in Fig.~\ref{fig:muI}, as first proposed in Ref.~\cite{Son:2000xc}.

Next let us consider $\SO_B$.
The Lagrangian of the theory is given by Eq.~(\ref{eq:Lagrangian}) 
with the color generators $T^a$ replaced by those of the gauge group $\SO(2N_c)$.  
A crucial difference from QCD is that there is no distinction  
between fermions in the fundamental and antifundamental representations
(in the vacuum) because the gauge group is real.  
For this reason, mesons in this theory are not necessarily neutral under $\U(1)_B$; 
baryon-number charged mesons (baryonic mesons or antibaryonic mesons) 
out of two quarks or two antiquarks also arise.  

In order to identify the lightest baryonic meson, 
let us consider the symmetry breaking pattern of the theory. 
When $m = \mu_B = 0$, 
the chiral symmetry, which at first sight looks $\SU(N_{f})_{L}\times \SU(N_{f})_{R} \times \U(1)_{B}$, 
is enhanced to $\SU(2N_f)$, where $N_f$ is the number of flavors.  
This enhancement of chiral symmetry originates from the fact that
there is no distinction between left- and right-handedness.
One can actually write the fermionic part of the Lagrangian 
manifestly invariant under $\SU(2N_f)$ using the new variable $\Psi=(\psi_L, \sigma_2 \psi_R^*)^T$:
\begin{eqnarray}
{\cal L}_{\rm f}=i \Psi^{\dag} \sigma_{\mu} D_{\mu} \Psi,
\end{eqnarray}
where $\sigma_{\mu}=(\sigma_k,-i{\bf 1}_2)$ with the Pauli matrices $\sigma_k$ ($k=1,2,3$).
The $\SU(2N_f)$ chiral symmetry is spontaneously broken to 
$\SO(2N_{f})$ by the formation of the chiral condensate 
$\langle \bar{\psi}{\psi} \rangle$, which gives rise to 
$2N_f^2 + N_f -1$ Nambu-Goldstone bosons parametrized by 
the coset space $\SU(2N_{f})/\SO(2N_{f})$: 
neutral mesons $\Pi_a=\bar{\psi} \gamma_{5} P_a \psi$, 
baryonic mesons (or simply diquark)
$\Sigma_S = \psi^{T} C \gamma_5 Q_S \psi$ and 
antibaryonic mesons (or antidiquark)
$\Sigma_S^{\dag}= \psi^{\dag} C \gamma_5 Q_S \psi^*$,
where $P_a$ are traceless and Hermitian $N_f \times N_f$ matrices, 
$P_a=P_a^{\dag}$ ($a=1,2,\cdots,N_f^2-1$), 
and $Q_S$ are symmetric $N_f \times N_f$ matrices, $Q_S^T=Q_S$
($S=1,2,\cdots,N_f(N_f + 1)/2$), in the flavor space.
In the vacuum, their masses are degenerate due to the unbroken flavor
symmetry, $m_\Pi=m_\Sigma=m_{\Sigma^\dagger}$. 
(At large $N_c$, because of the orbifold equivalence, their masses also 
coincide with the pion mass $m_\pi$ in QCD.)
This degeneracy is
resolved if we turn on nonzero $\mu$ which explicitly breaks 
the $\SO(2N_f)$ symmetry down to $\SU(N_f)_V \times \U(1)_B$.

At $T=0$, as we turn on $\mu_B$, the lightest particle with the baryon
number, the diquark $\Sigma_S$, condenses for $\mu > m_\Sigma/2$.
This is the BEC of diquarks where $\U(1)_B$ symmetry is broken to ${\mathbb Z}_2$. 
At sufficiently large $\mu_B$, on the other hand, the one-gluon exchange interaction 
between quarks near the Fermi surface is attractive in the color symmetric channel 
and gives rise to the pairing of the form 
$\langle \psi^{T} C \gamma_5 Q_S \psi \rangle$ \cite{Hanada:2011ju}.
This BCS pairing breaks the same symmetry as the BEC at small $\mu_B$, and 
it is again natural to expect the BEC-BCS crossover for 
$\mu>m_{\Sigma}/2$ along the $\mu$ axis.

In the same manner, one can also obtain the phase diagram of $\Sp_B$. 
(For further details, see Ref.~\cite{Hanada:2011ju}.) 

\section{Phase quenching in large-$N_c$ QCD}
\label{sec:large-Nc}
\subsection{Orbifold equivalence}\label{sec:orbifolding}
In this subsection, we briefly review the large-$N_c$ orbifold equivalence
\cite{Kachru:1998ys,Lawrence:1998ja,
Bershadsky:1998mb,Bershadsky:1998cb,Kovtun:2004bz}
and apply it to QCD and QCD-like theories 
\cite{Cherman:2010jj,Cherman:2011mh,Hanada:2011ju}. 
Thereby we establish the exactness of the phase quenching
in the large-$N_c$ limit (for details, see Ref.~\cite{Hanada:2011ju}).
The relations between QCD and QCD-like theories through the orbifold 
projections are summarized in Fig.~\ref{fig:equivalence}.

The idea of the orbifold equivalence is the following:
first we choose the discrete symmetry $P$ 
(subgroup of gauge, flavor, or spacetime symmetry)
of the original theory called the {\it parent}.
We then throw away all the degrees of freedom not invariant
under $P$. This procedure is called the 
{\it orbifold projection}. After the projection,
we obtain a new theory called the {\it daughter}. 
The orbifold equivalence states that, in the large-$N_c$ 't Hooft limit
where the 't Hooft coupling $g^2 N_c$ is kept finite,
correlation functions of operators ${\cal O}^{(p)}(A_{\mu},\psi)$
invariant under $P$ in the parent (called {\it neutral} operators)
agree with those of the operators 
${\cal O}^{(d)}(A_{\mu}^{\rm proj},\psi^{\rm proj})$
that consist of projected fields in the daughter: 
\beq
\langle{\cal O}_1^{(p)}{\cal O}_2^{(p)}\cdots\rangle_{p}
=
\langle{\cal O}_1^{(d)}{\cal O}_2^{(d)}\cdots\rangle_{d}.
\eeq
Here coupling constants should be appropriately related;
for example, for the equivalence between QCD$_B$ with $\SU(N_c)$ gauge group 
and $\SO_B$ with $\SO(2N_c)$ gauge group, which we shall consider below, 
we take
\begin{eqnarray}
g_{\SU}^2=g_{\SO}^2.
\end{eqnarray}
The field theoretic proof to all orders in the perturbation theory
was given by Bershadsky and Johansen \cite{Bershadsky:1998cb} and 
nonperturbative proof in certain gauge theories was
given by Kovtun, \"{U}nsal, and Yaffe \cite{Kovtun:2004bz}.
For QCD$_B$, QCD$_I$, SO$_B$, and Sp$_B$, a couple of evidences of nonperturbative equivalence were also provided by 
the weak-coupling analysis at high density limit \cite{Hanada:2011ju}, 
chiral perturbation theories \cite{Cherman:2011mh}, 
chiral random matrix models \cite{Hanada:2011ju}, and holographic models \cite{Hanada:2012nj}. 

Here let us take $\SO_B$ 
as a parent and consider the projections to QCD$_B$ and QCD$_I$
independently.\footnote{For an earlier work of the orbifold projection 
from $\SO(2N_{c})$ to $\SU(N_{c})$ gauge theories, see Ref.~\cite{Unsal:2006pj}. 
See also Ref.~\cite{Unsal:2007fb} where phase diagrams of 
QCD-like theories with different matter contents at small $S^3$ 
have been studied in the context of the orbifold equivalence.}
We identify the ${\mathbb Z}_4$ discrete symmetries of $\SO_B$  
generated by $J_c = -i\sigma_{2} \otimes 1_{N_{c}} \in \SO(2N_{c})$ 
and $\omega = e^{i \pi/2} \in \U(1)_B$, where 
$1_{N}$ is an $N \times N$ identity matrix.\footnote{Here $J_c$ is 
chosen to satisfy the {\it regularity condition}, 
${\rm Tr} (J_c^n) =0$ when $J_c^n \neq \pm 1_{2N_c}$.
This condition is necessary for the derivation of the orbifold equivalence 
\cite{Bershadsky:1998cb}.}
We require the gauge field $A^{\rm SO}_{\mu,ab}$ 
and the fermion $\psi^{\rm SO}_{\alpha,a}$ to be invariant under the following ${\mathbb Z}_2$ transformation 
embedded in the gauge and $\U(1)_B$ transformations \cite{Cherman:2010jj},
\begin{eqnarray}
\label{eq:gauge}
A^{\rm SO}_{\mu,ab} &=& (J_c)_{aa'} A^{\rm SO}_{\mu,a'b'} (J_c^{-1})_{b'b}, \\
\label{eq:fermion_baryon}
\psi^{\rm SO}_{\alpha,a} &=& \omega (J_c)_{aa'} \psi^{\rm SO}_{\alpha,a'}. 
\end{eqnarray} 
Under these projection conditions, QCD$_B$ is obtained as the daughter. 
In order to see it, we decompose the gauge field and fermion field 
of the parent SO$_B$ as
\begin{align}
A_\mu 
=i\left(
\begin{array}{cc}
A_\mu^A+B_\mu^A & C_\mu^A-D_\mu^S\\
C_\mu^A+D_\mu^S & A_\mu^A-B_\mu^A
\end{array}
\right), \qquad 
\psi=\left(
\begin{array}{c}
\xi+\zeta \\
i(\xi-\zeta)
\end{array}\right) ,
\end{align}
where the gauge fields with the superscript $A$ ($S$) 
are $N_{c}\times N_{c}$ anti-symmetric (symmetric) matrices
and $\xi$ and $\zeta$ are $N_c$-component fermions.  
Under the $\mathbb{Z}_{2}$ symmetry, $A_\mu^A$ and $D_\mu^S$ are even 
while $B_\mu^A$ and $C_\mu^A$ are odd, and 
the orbifold projection sets $B_{\mu}^{A} = C_{\mu}^{A} = 0$;
we can also see $\xi$ and $\zeta$ are even and odd under
$\mathbb{Z}_{2}$, respectively. 
Therefore, the daughter fields after the projection are  
\begin{align}
A_{\mu}^{\rm proj}
=
i\left(
\begin{array}{cc}
A_{\mu}^A  & -D_\mu^S\\
D_\mu^S & A_\mu^A
\end{array}
\right),
\qquad
\psi_f^{\rm proj}=\left(
\begin{array}{c}
\xi \\
i \xi
\end{array}\right)
\end{align}
After a unitary transformation using the matrix
\begin{eqnarray}
P = \frac{1}{\sqrt{2}}\left(
\begin{array}{cc}
1_{N_{c}} & i 1_{N_{c}} \\
1_{N_{c}} & -i 1_{N_{c}}
\end{array} 
\right),
\end{eqnarray}
it can be rewritten as 
\begin{eqnarray}
P A_{\mu}^{\rm proj} P^{-1} =
  \left(
\begin{array}{cc}
-({A}_{\mu}^{\U})^{T} & 0\\
0 & ({A}^{\U})_{\mu}
\end{array} 
\right),
\qquad 
P \psi_f^{\rm proj}
=\left(
\begin{array}{c}
0 \\
\psi^{\U}
\end{array}\right),
\nonumber \\
\end{eqnarray}
where ${A}_{\mu}^{\U} \equiv D_{\mu}^{S} + i A^{A}_{\mu}$ is 
a $\U(N_{c})$ gauge field and $\psi^{\U} = \sqrt{2} \xi$.  
Since the difference between $\U(N_{c})$ and $\SU(N_{c})$ is 
a $1/N_{c}^{2}$ correction and is negligible at large $N_c$,
the daughter theory can be regarded as QCD$_B$ given by
Eq.~(\ref{eq:Lagrangian}).
From this orbifold projection, we have the equivalence between
$\SO_B$ and QCD$_B$. However, the $\U(1)_B$ symmetry, whose
${\mathbb Z}_4$ subgroup is used for the projection of the fermion
in Eq.~(\ref{eq:fermion_baryon}), is spontaneously broken to 
${\mathbb Z}_2$ in the diquark condensation phase
(the BEC/BCS region in Fig.~\ref{fig:SO}); the equivalence
is valid only outside that region.

One can also construct the projection from $\SO_B$  
to QCD$_I$ by choosing another ${\mathbb Z}_2$ symmetry 
\cite{Cherman:2010jj,Hanada:2011ju},
\begin{eqnarray}
A^{\rm SO}_{\mu,ab} &=& (J_c)_{aa'} A^{\rm SO}_{\mu,a'b'} (J_c^{-1})_{b'b}, \\
\label{eq:fermion_isospin}
\psi_{\alpha,af}^{\rm SO} &=& (J_c)_{aa'} \psi_{\alpha,a'f'}^{\rm SO} (J_{i}^{-1})_{f'f},
\end{eqnarray}
where $J_{i} = - i\sigma_2 \otimes 1_{N_f/2}$ generates ${\mathbb Z}_4$ subgroup
of ${\rm SU}(2)$ isospin symmetry and the projection condition 
for the gauge field is the same as Eq.~(\ref{eq:gauge}). 
To see how QCD$_I$ can be obtained through the projection, 
we decompose the flavor $2N_f$-component fundamental fermion into 
two $N_f$-component fields, 
\begin{eqnarray}
\psi^{\rm SO}
=(\psi_i \ \psi_j), 
\end{eqnarray}
with $i$ and $j$ being the isospin indices, 
and furthermore decompose $2N_c$ color components to two sets of $N_c$ components, 
\begin{eqnarray}
\psi_i 
=
\left(
\begin{array}{c}
\xi_i\\
\zeta_i
\end{array}
\right),
\qquad
\psi_j 
=
\left(
\begin{array}{c}
\xi_j\\
\zeta_j
\end{array}
\right). 
\end{eqnarray}
If we define $\psi_{\pm}=(\xi \pm i \zeta)/\sqrt{2}$,
$\varphi_{\pm}=(\psi_{\pm}^i \mp i \psi_{\pm}^j)/\sqrt{2}$
and $\chi_{\pm}=(\psi_{\pm}^i \pm \ i \psi_{\pm}^j)/\sqrt{2}$,
the fermions $\varphi_{\pm}$ survive but $\chi_{\pm}$ disappear after the projection
(\ref{eq:fermion_isospin}).
Because $\varphi_{\pm}$ couple to $(A_{\mu}^{\rm SU})^C$ and $A_{\mu}^{\rm SU}$ respectively, 
the Lagrangian of the daughter theory is now
\begin{eqnarray}
{\cal L}_{{\rm QCD}_I} &=& \frac{1}{4 g_{\rm SU}^{2} } \tr ({F}^{\SU}_{\mu \nu})^2
\nonumber \\
& &+ \sum_{f, \pm} \bar{\psi}^{\SU}_{f \pm}\left( \gamma^{\mu} {D}_{\mu} 
+ m \pm \mu \gamma^4 \right)\psi^{\SU}_{f \pm},
\end{eqnarray}
where $\psi^{\rm SU}_{+}=\sqrt{2}\varphi_-$ and $\psi^{\rm SU}_{-}=\sqrt{2}\varphi_+^C$.
This theory is QCD$_I$.
In this case, the isospin symmetry used for the projection of the fermion 
is unbroken everywhere, and so the orbifold equivalence holds 
including the BEC/BCS region of the phase diagram.
Therefore, through the equivalence with $\SO_B$, 
we obtain the equivalence between QCD$_B$ and QCD$_I$ 
outside the BEC/BCS region of the latter; 
the phase quenching is exact for neutral sectors in this region.
The same conclusion can be reached through the equivalence with
$\Sp_B$ \cite{Hanada:2011ju}.

A few remarks are in order here.  
Firstly, not all the operators coincide. 
In the parent theory, only the operators invariant under the projection symmetry $P$ 
are related to the counterparts in the daughter. 
In the daughter, not all the operators are obtained from the parent through 
the projections. 
As an example, consider the fate of neutral pions and (anti)diquarks 
of the parent $\SO_B$ after the orbifold projections (see Tab.~\ref{tab:example}). 
The projection to QCD$_B$ maps $\U(1)_B$-neutral pions of $\SO_B$  
to pions in QCD$_B$, $\pi^0, \pi^+$ and $\pi^-$, and throws away (anti)diquarks. 
On the other hand, the projection to QCD$_I$ sends 
(isospin singlet part of) $\U(1)_B$-neutral pions, diquarks, antidiquarks, to $\pi^0$, $\pi^+$ and $\pi^-$, respectively. 
Therefore the diquarks in $\SO(2N_c)$ theory and $\pi^+$ in QCD$_I$
have the same mass $m_{\pi}$ in the vacuum and 
the same excitation energy at any $\mu$ (at $T=0$). 
In the same way as $\pi^+$ condenses in QCD$_I$ at $\mu=m_\pi/2$, 
the diquark condenses at $\mu=m_\pi/2$ (see Fig.~\ref{fig:muI} and Fig.~\ref{fig:SO}). 
Note that the charged pions $\pi^\pm$ in QCD$_B$ and QCD$_I$ have different origins in SO$_B$ 
and they do not correspond each other.

\begin{table}[b]
\caption{Some examples of the correspondence between $\SO_B$, QCD$_I$ and QCD$_B$
through the orbifold projections.
See the text for detail.}
\label{tab:example}
\begin{tabular}{|c|c|c|c|c|}
\hline
    & Order parameter & \multicolumn{3}{|c|}{Elementary excitations} \\ \hline \hline
$\SO_B$  & $\langle \bar \psi \psi \rangle_{\SO_B}$ & Neutral pions & Diquarks & Antidiquarks \\
QCD$_I$ & $\langle \bar \psi \psi \rangle_{{\rm QCD}_I}$ & $\pi^0$ & $\pi^+$ & $\pi^-$ \\
QCD$_B$ & $\langle \bar \psi \psi \rangle_{{\rm QCD}_B}$ & $\pi^0, \pi^+, \pi^-$ & $\times$ & $\times$ \\
\hline
\end{tabular}
 \end{table}

Secondly, note that two projections \eqref{eq:fermion_baryon} and 
\eqref{eq:fermion_isospin} are equivalent when $\mu=0$ as they should be. 
Both are a $\mathbb{Z}_4$ subgroup of the flavor symmetry which 
mixes two Majorana flavors. 
Once $\mu$ ($\mu_B$ or $\mu_I$) is turned on, they are not equivalent. 
The flavor symmetry $J_i$ used in \eqref{eq:fermion_isospin} is 
essentially the same as $J_c$, and the proof in Ref.~\cite{Bershadsky:1998cb} 
can be repeated straightforwardly \cite{Hanada:2011ju}; 
the only difference is some color-index loops 
which are replaced by flavor-index loops. 
On the other hand, $\mathbb{Z}_4\in \U(1)_B$ used in Eq.~\eqref{eq:fermion_baryon} 
is different and the proof in Ref.~\cite{Bershadsky:1998cb} holds only 
for planar diagrams with at most one fermion loop.  
Because fermion loops are suppressed by the factor $N_f/N_c$, 
the equivalence through the projection \eqref{eq:fermion_baryon} to QCD$_B$  
holds in the 't Hooft large-$N_c$ limit ($N_c \rightarrow \infty$ with $N_f$ fixed) while the one 
through the projection to QCD$_I$ \eqref{eq:fermion_isospin} holds 
also in the Veneziano large-$N_c$ limit ($N_c \rightarrow \infty$ with $N_f/N_c$ fixed)
\cite{Hanada:2011ju}. 

The above second remark has an implication for the $1/N_c$ corrections
\cite{Hanada:2011ju}. 
Compare QCD$_B$ and QCD$_I$. In the 't Hooft large-$N_c$ limit, 
expectation values of gluonic operators trivially agree because 
the fermions are not dynamical.
Now consider finite-$N_c$, say $N_c=3$ and $N_f=2$. 
Then the largest correction to the 't Hooft limit comes from 
one-fermion-loop planar diagrams, which, as we have seen, 
do not distinguish $\mu_B$ and $\mu_I$. 
Therefore the difference of expectation values of gluonic operators 
is at most $(N_f/N_c)^2$ (two-fermion-loop planar diagrams). 
In particular, the deconfinement temperatures, which are determined by 
the Polyakov loop, agree up to corrections of this order. A similar observation was made 
in Ref.~\cite{Toublan:2005rq} by a perturbative argument. 

Note that the $1/N_c$ correction can become larger in the confining phase, 
because of thermal excitations of pions, resonances, and baryons, 
which large-$N_c$ arguments do not take into account; 
baryon gas in QCD$_B$ is quite different from pion gas in QCD$_I$ \cite{Han:2008xj}. 
On the other hand, for $T>T_c$, fundamental degrees of freedom are
deconfined quarks and gluons rather than baryons and mesons, 
where the difference between QCD$_B$ and QCD$_I$ becomes much smaller 
and the large-$N_c$ equivalence may be well satisfied even at $N_c=3$.
It can indeed be confirmed numerically, as we will see in Sec.~\ref{sec:lattice}. 
The results there show that the phase quenching 
is a very useful tool for the study of the chiral transition.  

\subsection{Implications for the phase reweighting}
\label{sec:overlap}
In the phase reweighting method, one calculates observables (e.g., the chiral condensate) 
by using the QCD$_I$ ensemble and by taking into account the phase factor.
There, the absence of a severe overlap problem is implicitly assumed. 
But how can one justify this assumption?
Actually we can obtain a very simple answer by combining the orbifold equivalence 
and a simple symmetry argument:  
any observable coincide to $O(N_f/N_c)$
implying that there is no overlap problem for these quantities to this order.

At first it may sounds implausible; for example, the baryon number density takes nonzero 
value in QCD$_B$,  while it is zero in QCD$_I$, but then how the phase quenching can work?  

The point is that, although the partition functions of
QCD$_I$ and the phase-quenched QCD are equivalent,
observables are not necessarily the same. For example, in QCD$_I$, 
the propagators of up and down quarks are $D^{-1}(+\mu)$
and $D^{-1}(-\mu)$, while in the phase-quenched QCD, 
both of them are $D^{-1}(+\mu)$.
(In the terminology of the lattice simulation, we generate the
configuration by using QCD$_I$, but the measurement is done
by using the same operators as QCD$_B$.)
As a result, the expressions for the chiral condensate and the 
baryon/isospin density in each theory are given by
\begin{widetext}
\begin{eqnarray}
\begin{array}{|c|c|c|c|}
\hline
& {\rm QCD}_B & {\rm QCD}_I & {\rm Phase\ quenched\ QCD}  \\
\hline
\langle \bar \psi \psi \rangle &2 \langle \tr D^{-1}(\mu)\rangle_B
&  \langle \tr D^{-1}(\mu) + \tr D^{-1}(-\mu)\rangle_I
& 2 \langle \tr D^{-1}(\mu)\rangle_I\\
\hline
\langle n_B \rangle &2 \langle \tr \gamma^0 D^{-1}(\mu)\rangle_B
&  0
&  2\langle \tr \gamma^0D^{-1}(\mu)\rangle_I \\
\hline
\langle n_I \rangle & 0
&  \langle \tr \gamma^0D^{-1}(\mu)- \tr \gamma^0D^{-1}(-\mu)\rangle_I
&  0\\
\hline
\end{array}
\nonumber
\end{eqnarray}
\end{widetext}
Because $\langle \tr D^{-1}(\mu)\rangle_I = \langle \tr D^{-1}(-\mu)\rangle_I$
due to the charge conjugation invariance, the chiral condensate 
in QCD$_I$ and that in the phase-quenched QCD take the same value. 
Therefore, the orbifold equivalence states that it 
remains unchanged by the phase quenching.
For the baryon density, we first note that $\langle n_I \rangle_I = \langle n_B \rangle_I$ 
because $\langle \tr \gamma^0D^{-1}(-\mu)\rangle_I
=- \langle \tr \gamma^0D^{-1}(\mu)\rangle_I$.
By combining it with the orbifold equivalence, $\langle n_B \rangle_B = \langle n_I \rangle_I$,
we conclude that the phase quenching is valid for the baryon density. 
Note also that the isospin density is trivially zero in QCD$_B$ and 
in the phase-quenched QCD 
it is $\langle \tr \gamma^0D^{-1}(\mu) - \tr \gamma^0D^{-1}(\mu) \rangle_I =0$,
and the phase quenching is valid. 
The same argument is applicable to other fermionic observables too.

The exactness of the phase quenching in the large-$N_c$ limit 
can be rephrased that the overlap problem is $1/N_c$-suppressed.  
Note however that the phase reweighting is doable only at small volume and $N_c$;
the average phase factor is exponentially suppressed at large volume and/or $N_c$.
\subsection{Relation to the quenched approximation}\label{sec:quenched_approximation} 
Let us consider the quenched approximation in lattice QCD, 
in which the fermionic observables are calculated by using the gauge configurations generated 
in the quenched QCD (i.e., the pure Yang-Mills theory). 
This approximation is believed to become exact in the 't Hooft large-$N_c$ limit 
where dynamical fermion loops are suppressed. 
Does this approximation make sense at finite $\mu$? 

For concreteness, let us consider the chiral condensate. 
It is calculated as 
\begin{eqnarray}
\left\langle\bar{\psi}_f\psi_f\right\rangle_{\rm YM}
=
\left\langle \tr\left(D^{-1}_f(A,\mu_f)\right)\right\rangle_{\rm YM}, 
\end{eqnarray} 
where the propagator $D^{-1}_f(A,\mu)$ is a function of the gauge field $A_\mu$ 
and the path integral is taken by using the pure Yang-Mills action. As long as one considers 
a perturbation around the trivial vacuum, it contains all one-fermion-loop diagrams and  
gives the right answer at large $N_c$.  However whether it is correct at nonperturbative level 
is nontrivial, and actually it fails at finite $\mu$ -- 
although QCD$_B$ and QCD$_I$ are identical in the quenched approximation
as we have seen in Sec.~\ref{sec:introduction}, QCD$_B$ and QCD$_I$ actually have completely different phase diagrams. 
Then what is wrong with the quenched approximation? 

The expectation values in QCD$_B$ and QCD$_I$ are written in terms of the quenched QCD as 
 \begin{eqnarray}
& &
\left\langle\bar{\psi}_f\psi_f\right\rangle_{B,I}
=
\nonumber\\
& &
\frac{\left\langle \tr\left(D^{-1}_f(A,\mu_f)\right)\cdot \prod_f
\det D_f(A,\mu_f)\right\rangle_{\rm YM}}{
\left\langle \prod_f\det D_f(A,\mu_f)\right\rangle_{\rm YM}
}.  
\end{eqnarray} 
They agree with the value under the quenched approximation 
when the following factorization holds, 
\begin{eqnarray}
& &
\left\langle \tr\left(D^{-1}_f(A,\mu_f)\right)\cdot \prod_f\det D_f(A,\mu_f)\right\rangle_{\rm YM}
\nonumber\\
&=&
\left\langle \tr\left(D^{-1}_f(A,\mu_f)\right)\right\rangle_{\rm YM}
\cdot 
\left\langle \prod_f\det D_f(A,\mu_f)\right\rangle_{\rm YM}. 
\nonumber\\
\end{eqnarray} 
This factorization should be distinguished from the usual one
which follows from the 't Hooft counting; although a finite number 
of traces factorize, the factorization is not valid for determinants.
It is plausible that the factorization takes place in QCD$_I$, 
because the quenched approximation exhibits the pion condensation,
as is explicitly demonstrated within the chiral random matrix model 
\cite{Stephanov:1996ki}.
However, the factorization obviously fails in QCD$_B$
for $\mu > m_\pi/2$ at $T=0$ where the pion condensation should not occur.
Although we have not proven in this paper, the equivalence between QCD$_{B,I}$ 
and the quenched QCD outside the pion condensation region at large $N_c$
would be useful because the quenched QCD is numerically cheaper.
It is also interesting to calculate the chiral condensate and the baryon/isospin density 
by using lattice configurations with dynamical fermions at $\mu=0$.  
  
\section{Phase quenching in effective models of QCD}
\label{sec:model}
As we have seen, the phase quenching is exact in the large-$N_c$ limit 
for a class of observables outside the pion condensation phase of 
the phase-quenched theory (QCD$_I$). 
Therefore effective models of QCD, if describe underlying QCD properly, 
should exhibit the same large-$N_c$ equivalence. 
Then the phase quenching is expected to be exact 
in the mean-field approximation (MFA) of the models, 
since MFA corresponds to the leading-order in the $1/N_c$ expansion of QCD. 

In this section, we show this statement by explicitly constructing 
orbifold projections within several models frequently used to 
study the phase diagram of QCD, 
including the Nambu--Jona-Lasinio (NJL) model, linear sigma model (L$\sigma$M), 
Polyakov--Nambu--Jona-Lasinio (PNJL) model, Polyakov-quark-meson (PQM) model, 
chiral random matrix model ($\chi$RMM), and 
strong-coupling expansion of lattice QCD.
(The orbifold projections of the $\chi$RMM were already 
given in Ref.~\cite{Hanada:2011ju}, but we include it here for completeness.)
We then analytically demonstrate the exactness of the phase quenching in 
NJL model \cite{Toublan:2003tt} and $\chi$RMM \cite{Klein:2003fy, Hanada:2011ju}, 
for which the effective potentials are known. 

\subsection{Mean-field approximation}
\label{sec:MFA}
\begin{figure}[t]
\begin{center}
\includegraphics*[width=5cm]{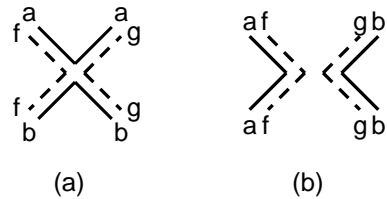}
\end{center}
\vspace{-0.5cm}
\caption{Two types of interactions: (a) $(\bar \psi_{af} \psi_{ag})(\bar \psi_{bg} \psi_{bf})$ 
originating from the one-gluon exchange interaction and (b) 
$(\bar \psi_{af} \psi_{af})(\bar \psi_{bg} \psi_{bg})$ originating 
from the instanton-induced interaction, with $a,b$ ($f,g$) being color (flavor) indices.
The solid and dotted lines denote color and flavor lines, respectively.}
\label{fig:interaction}
\end{figure}

\begin{figure}[t]
\begin{center}
\includegraphics*[width=7.5cm]{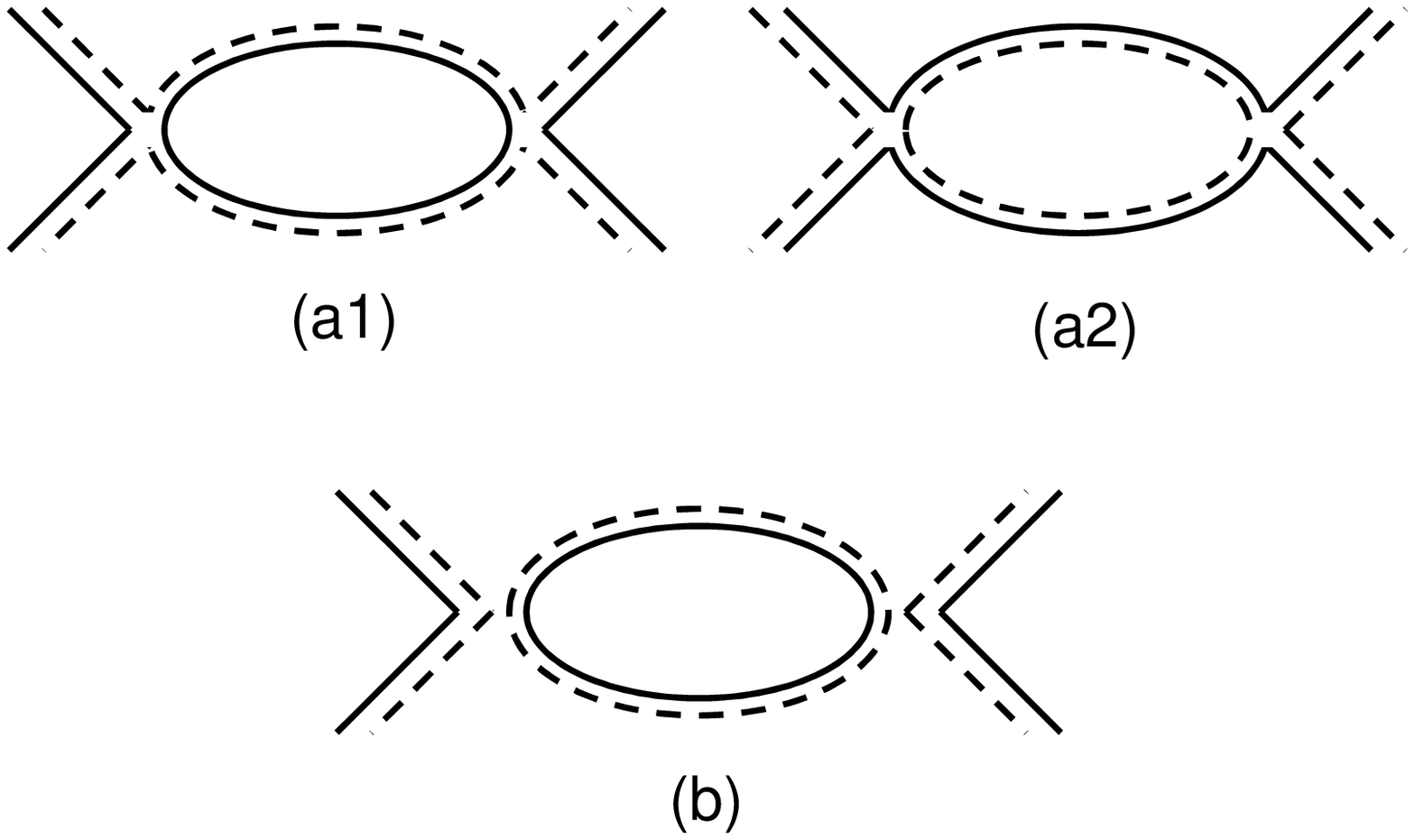}
\end{center}
\vspace{-0.5cm}
\caption{Loop diagrams induced by the two types of interactions (a) and (b)
in Fig.~\ref{fig:interaction}: (a1) $O(N_c^{-1})$ with no flavor loop, 
(a2) $O(N_c^{-2})$ with one flavor loop, 
and (b) $O(N_c^{-1})$ with one flavor loop.}
\label{fig:loop}
\end{figure}

In order to apply the large-$N_c$ equivalence to the effective models,
let us set up the $1/N_c$-counting scheme in the models, 
so that the right powers of $1/N_c$ in QCD are reproduced. 
As an example, we consider the NJL model (see Sec.~\ref{sec:njl} for the
Lagrangian). The quark field $\psi$ has $N_c$ colors so that a closed color loop gives a factor of $N_c$. 
(Here $N_c$ is treated as a variable and will be taken to $N_c=3$ at the end of calculations.)
We need to make the counting scheme such that each flavor loop gives 
a suppression factor of $1/N_c$. 
Then the coupling constant of the four-fermi interaction should be taken as $O(N_c^{-1})$, 
and furthermore, 
the form of possible four-fermi interactions are restricted:
the interaction of the form 
$(\bar \psi_{af} \psi_{ag})(\bar \psi_{bg} \psi_{bf})$
shown in Fig.~\ref{fig:interaction}(a) is allowed,
but $(\bar \psi_{af} \psi_{af})(\bar \psi_{bg} \psi_{bg})$
in Fig.~\ref{fig:interaction}(b) is not,
where $a,b$ ($f,g$) are color (flavor) indices. This is because
the one-flavor-loop diagram in Fig.~\ref{fig:loop}(b)
derived from the latter interaction is not suppressed in $1/N_c$ compared
with the diagram with no flavor loop in Fig.~\ref{fig:loop} (a1)
derived from the former. 
Once we exclude the interaction of the form $(\bar \psi_{af} \psi_{af})(\bar \psi_{bg} \psi_{bg})$, 
the right $1/N_c$-counting follows and we can use the same proof of 
the orbifold equivalence in the NJL model as the large-$N_c$ QCD in Ref.~\cite{Hanada:2011ju}.

Under these conditions, we now recall the strategy to compute the 
effective potential of the NJL model.
We first perform the Hubbard-Stratonovich transformation
by introducing auxiliary fields corresponding to the fermion bilinears,
$\sigma_A = (G/N_c)\bar \psi \tau_A \psi$ and 
$\pi_A = (G/N_c)\bar \psi i \gamma_5 \tau_A \psi$,
where $\tau_A$ are the $\U(2)$ flavor generators.
We then integrate out fermions to obtain the partition function
\beq
Z \equiv e^{-W} = \int d \sigma_A d \pi_A e^{-I(\sigma_A, \pi_A)}.
\eeq
Here $I(\sigma, \pi)$ is the bosonized action 
\beq
I(\sigma_A, \pi_A) = N_c \left[-\tr \log D + \frac{1}{G}\int d^4 x (\sigma_A^2 + \pi_A^2)\right],
\nonumber \\
\eeq
with $D= \gamma^{\mu} \partial_{\mu} + 2(\sigma_A + \pi_A)$. 
This effective action describes a theory for bosonic 
fields (mesons) $\sigma_A$ and $\pi_A$.
Because there is an overall factor $N_c$ in the action, the expansion 
of $Z$ [or $I(\sigma_A, \pi_A)$] in terms of $1/N_c$ is equivalent to the expansion 
in terms of meson loops \cite{Coleman:1973jx}. In particular, the leading 
order in $1/N_c$ corresponds to the saddle-point approximation, 
or the MFA where the auxiliary fields are 
replaced by the expectation values (i.e., the mean fields).\footnote{A 
similar discussion on the $1/N$ expansion in the context of 
condensed matter physics, e.g., fermions at unitarity, can be found in 
Refs.~\cite{Nikolic:2007zz, Abuki:2008tj} where $N$ is the number of species of fermions. 
In Ref.~\cite{Nikolic:2007zz} $1/N$ corrections are found to be numerically 
small by Monte Carlo simulations.}
In the language of many-body physics, 
the MFA corresponds to the Hartree approximation 
for the quark self-energy and to the random phase approximation
for the four-fermi interaction (see, e.g., Ref.~\cite{Fetter}).
In order to go beyond the MFA, we have to take into account
$n$-meson-loops ($n \in {\mathbb N}$) order by order which
give $1/N_c^n$ corrections to the MFA \cite{Dmitrasinovic:1995cb}.

Since the L$\sigma$M, PNJL model, and PQM model can be 
regarded as simple generalizations of the NJL model, 
as we will argue below, we can develop similar counting schemes 
to apply the orbifold equivalence to them. 
In the case of the $\chi$RMM, we can use the 
large-$N$ equivalence by identifying the the size of the matrix $N$ 
as a variable \cite{Hanada:2011ju}, which is finally taken to be infinity 
corresponding to the thermodynamic limit.

In the following, we will argue the orbifold equivalence in each model. 
It then predicts the exactness of the phase quenching in 
the MFA outside the pion condensation phase of the phase-quenched model 
(which we denote ``model$_I$").

\subsection{Nambu--Jona-Lasinio model}
\label{sec:njl}
We first consider the Nambu--Jona-Lasinio (NJL) model \cite{Nambu:1961tp}
which captures the physics of chiral symmetry breaking in QCD
(for reviews, see Refs.~\cite{Hatsuda:1994pi, Buballa:2003qv}).
In order to simplify the discussion, we consider the chiral limit 
$m=0$ so that we maintain the full chiral symmetry of the model. 
The generalizations to include nonzero $m$ is straightforward.

The starting point is the Lagrangian with the $\U(N_c)$ color current 
interaction with $N_f$ flavors,
\begin{eqnarray}
\label{eq:NJL_current}
{\cal L}_{\rm NJL}
=\bar{\psi}_f\left(
\gamma^\mu\partial_\mu + \mu_f\gamma^4
\right)\psi_f
- \frac{G}{N_c} J^{(\U)}_{\mu A} J^{(\U)}_{\mu A}, 
\end{eqnarray}
where $J_{\mu A}^{({\U})} = \bar{\psi}_f\gamma_\mu T_{\U}^A\psi_f$ and 
$T^A_{\U}$ are the $\U(N_c)$ color generators and 
summation is taken over repeated indices. 
The coupling constant $G$ is taken to be of order $N_c^{0}$.
One rewrites it keeping only the interactions in the 
scalar and pseudoscalar channels after Fierz transformations:
\beq
{\cal L}_{\rm NJL}
\label{eq:NJL}
&=&
\bar{\psi} \left(\gamma^\mu \partial_\mu + \mu_f\gamma^4 \right) \psi + {\cal L}_{\rm int},
\\
\label{eq:4-fermi}
{\cal L}_{\rm int}&=& 
- \frac{G}{N_c} \left[
(\bar{\psi}_f\psi_{f'})(\bar{\psi}_{f'}\psi_{f}) + 
(\bar{\psi}_f i\gamma^5\psi_{f'})(\bar{\psi}_{f'} i\gamma^5\psi_{f})
\right]. 
\nonumber \\
\eeq
In the Lagrangians (\ref{eq:NJL_current}) and (\ref{eq:NJL}), 
the invariance under $\U(N_c)$ gauge symmetry 
and $\U(N_f)_L \times \U(N_f)_R$ flavor symmetry are manifest.
Here we ignore the effect of instantons or the $\U(1)_A$ anomaly
which explicitly breaks the $\U(1)_A$ symmetry, because it is
subleading in $1/N_c$;\footnote{To understand this statement, let us compare 
the couplings of the one-gluon exchange interaction $G_{\rm OGE}$ 
and the instanton-induced interaction $G_{\rm inst}$, where $G_{\rm OGE}$
and $G_{\rm inst}$ are defined as the coefficients of multi-fermi interactions
[see Eq.~(\ref{eq:inst}) for the form of the instanton-induced interaction]. 
One then finds that
$G_{\rm OGE} \sim g^2 \sim N_c^{-1}$ in the 't Hooft limit 
(where $g$ is the QCD coupling constant),
as is consistent with the $1/N_c$ counting for $G/N_c \sim N_c^{-1}$ 
in Eq.~(\ref{eq:NJL_current}). On the other hand, 
$G_{\rm inst} \sim N_c^{-N_f}$, since $G_{\rm inst}$ is related to the 
$\eta'$ meson mass as
\beq
m_{\eta'}^2 \sim G_{\rm inst}\frac{\sigma^{N_f}}{f_{\eta'}^2} \sim N_c^{-1},
\eeq
where $\sigma \sim N_c$ is the chiral condensate, $f_{\eta'} \sim N_c^{1/2}$
is the decay constant of $\eta'$, and the relation $m_{\eta'}^2 \sim N_c^{-1}$ 
follows from the Witten-Veneziano formula. 
Therefore, the instanton-induced interaction
is suppressed compared with the one-gluon exchange interaction for $N_f \geq 2$
in the $1/N_c$ counting.} 
from the viewpoint of the orbifold equivalence, 
there is no reason for the exactness of the phase quenching
if we take into account the $1/N_c$-suppressed instanton effects.
However, as we shall see in Sec.~\ref{sec:instanton}, even if we incorporate them, 
the phase quenching for the chiral condensate turns out to be exact within 
the NJL model at the level of MFA. 

For $\SO(2N_c)$ theory, we can construct the corresponding 
NJL model in the same manner, by starting with 
\begin{eqnarray}
\label{eq:SO-NJL_current}
{\cal L}
= i \Psi^\dagger_{F}\left(\sigma^\mu\partial_\mu + \mu_F\sigma^4 \right)\Psi_F
- \frac{G}{N_c} J^{(\SO)}_{\mu A} J^{(\SO)}_{\mu A}, 
\end{eqnarray}
where 
\beq
\label{eq:large_fermion}
\Psi_F = \left(\psi_{fL}, \psi_{fR}^c \right)
= \left(\psi_{fL}, C(\bar{\psi}_{fR})^T \right),
\eeq
$(F=1,2,\cdots,2N_f)$ are $2N_c$ component fermions and
the current $J^{(\SO)}$ is defined by 
\beq
J_{\mu A}^{(\SO)}=\bar{\psi}_f\gamma_\mu T^A_{\SO}\psi_f
=\bar{\Psi}_F\sigma_\mu T^A_{\SO}\Psi_F.
\eeq 
The invariance under $\SO(2N_c)$ gauge transformation and 
$\SU(2N_f)$ flavor rotation is manifest at this level. 
After the Fierz transformations and concentrating on the
interactions in the scalar and pseudoscalar channels, 
Lagrangian reduces to
\begin{eqnarray}
\label{eq:SO-NJL}
{\cal L}_{\rm NJL}^{\SO}
&=& i \Psi^\dagger_{F} \left(\sigma^\mu\partial_\mu + \mu_F\sigma^4 \right) \Psi_F
\nonumber \\
\! \!  & &
- \frac{G}{N_c} \bigl[
(\bar{\psi}_f\psi_{f'})(\bar{\psi}_{f'}\psi_{f})
+
(\bar{\psi}_f i \gamma^5\psi_{f'})(\bar{\psi}_{f'} i \gamma^5\psi_{f})
\nonumber \\
& &
+ (\bar{\psi}_fC\bar{\psi}^T_{f'})(\psi_{f'}^TC\psi_{f})
+ (\bar{\psi}_f i \gamma^5 C\bar{\psi}^T_{f'})(\psi^T_{f'} i \gamma^5 C\psi_{f})
\bigl]. 
\nonumber \\
\end{eqnarray}
The baryon chemical potential in $\SO(2N_c)$ theory corresponds to
$\mu_{F}=(\mu_{fL},-\mu_{fR})=(+\mu,-\mu)$. Note the minus sign in 
front of $\mu_{fR}$, which arises because of the charge conjugation. 
Due to this sign, the chiral symmetry is explicitly broken to 
$\SU(N_f)_L \times \SU(N_f)_R$. 

From the $\SO(2N_c)$ NJL model, it is possible to get 
$\SU(N_c)$ NJL model at finite $\mu_B$ (NJL$_B$) 
and $\SU(N_c)$ NJL model at finite $\mu_I$ (NJL$_I$). 
The projections are the same as those used for the fermion for
$\SO(2N_c)$ gauge theory to QCD$_B$ or QCD$_I$, 
(\ref{eq:fermion_baryon}) and (\ref{eq:fermion_isospin}):
\beq
\psi^{\rm SO}_{a} &=& \omega (J_c)_{aa'} \psi^{\rm SO}_{a'},
\nonumber \\
\psi_{af}^{\rm SO} &=& (J_c)_{aa'} \psi_{a'f'}^{\rm SO} (J_{i}^{-1})_{f'f},
\nonumber
\eeq
respectively.
Because the fermion kinetic term of the NJL model is the same as that of
$\SO(2N_c)$ gauge theory except that the gauge field is now absent, 
the projection conditions above lead to the fermion kinetic term 
with $\mu_B$ or $\mu_I$ (without the gauge field), respectively.

The projection of the four-fermi interaction is also simple.
Since the the current interaction of $\SO(2N_c)$ NJL model is
mapped to the current interaction of $\SU(N_c)$ NJL model,
if we concentrate on scalar and pseudoscalar sectors and
eliminate all of the others, the resultant interactions in Eqs.~(\ref{eq:SO-NJL}) 
and (\ref{eq:NJL}) must correspond to each other. 
From these orbifold projections, NJL$_B$ and NJL$_I$ are 
equivalent in the MFA outside the pion condensation phase of the latter. 

In order to check the exactness of the phase quenching, 
let us look at the free energy of two-flavor NJL model 
with quark chemical potentials $\mu_u$ and $\mu_d$. 
This was already computed in Ref.~\cite{Toublan:2003tt}.
In the absence of the pion condensation, the free energy 
in the MFA is given by 
\begin{align}
\label{eq:NJL_energy1}
\Omega_{\rm NJL}(& \mu_u, \mu_d, T) 
\nonumber \\
= & -\frac{N_c}{\pi^2}\int dp \ p^2 
\sum_{\pm, f={\rm u,d}} \left[ E_f + T \ln (1 + e^{-(E_f \pm \mu_f )/T}) \right]
\nonumber \\
&+ \frac{2G}{N_c}( \sigma_u^2 + \sigma_d^2 ),
\end{align}
where $E_f = \sqrt{p^2 + M_f^2}$ and $M_f = m_f - (4G/N_c) \sigma_f$.
From the expression above, the effective potential satisfies the relation
\beq
\label{eq:property}
\Omega_{\rm NJL}(\mu_B)|_{\mu_I=0}
=\Omega_{\rm NJL}(\mu_I)|_{\mu_B=0}.
\eeq
Thus, the chiral condensates $\sigma_f$ obtained from the gap equation 
\beq
\frac{\partial \Omega_{\rm NJL}}{\partial \sigma_f}=0,
\qquad (f={\rm u,d})
\eeq
coincide, 
\beq
\label{eq:equivalence_chiral}
\sigma_f(\mu_B)|_{\mu_I=0}
=\sigma_f(\mu_I)|_{\mu_B=0}.
\eeq
Therefore, the phase quenching for the free energy and
chiral condensate is exact in the MFA.

\subsubsection*{Instanton-induced interaction}
\label{sec:instanton}
The instanton-induced interaction, also known as the 
Kobayashi-Maskawa-'t Hooft interaction \cite{Kobayashi:1970ji},
is used in the practical calculations of the NJL model 
at the level of the MFA (for the treatment of the instanton-induced 
interaction in the MFA, see, e.g., Ref.~\cite{Hatsuda:1994pi}).
The instanton-induced interaction is suppressed compared with the 
one-gluon exchange interaction in the $1/N_c$ counting 
(see the footnote in Sec.~\ref{sec:MFA}).
However, even if we include this interaction,
the phase quenching is still exact.
This is because contributions of different chemical 
potentials, $\mu_u$ and $\mu_d$, decouple in the MFA, 
independently of the forms of interactions, 
as we shall show explicitly below [see Eq.~(\ref{eq:NJL_energy2})].

The instanton-induced interaction has the form
\beq
\label{eq:inst}
{\cal L}_{\rm inst} = -G_{\rm inst} 
\det_{f,g} \left[\bar \psi_f (1 + \gamma_5) \psi_g \right] + {\rm h.c.},
\eeq
which respects $\SU(N_f)_L \times \SU(N_f)_R \times \U(1)_B$, 
but breaks $\U(1)_A$ explicitly.
For two flavors, this can also be rewritten as
\begin{align}
{\cal L}_{\rm inst} = &-\frac{G_{\rm inst}}{2}[(\bar \psi \psi)^2 - 
(\bar \psi \tau^a \psi)^2 - (\bar \psi i \gamma_5 \psi)^2 \nonumber \\
&+ (\bar \psi i \gamma_5 \tau^a \psi)^2],
\end{align}
where $\tau^a$ are the $\SU(2)$ flavor generators.
The free energy in the MFA is computed as \cite{Toublan:2003tt}
\beq
\label{eq:NJL_energy2}
\Omega_{\rm NJL + inst}(\mu_u, \mu_d, T) = \Omega_{\rm NJL}(\mu_u, \mu_d, T) 
+ 2G_{\rm inst} \sigma_u \sigma_d,
\nonumber \\
\eeq
where $\Omega_{\rm NJL}(\mu_u, \mu_d, T)$ is given by (\ref{eq:NJL_energy1})
with $E_f$ replaced by $\tilde E_f = \sqrt{p^2 + \tilde M_f^2}$, and
$\tilde M_f = m_f - (4G/N_c) \sigma_f - 2G_{\rm inst} \sigma_{f'}$ with 
$f' \neq f$.

Despite the presence of mixing terms, such as $\sim \sigma_u \sigma_d$,
one finds the potential (\ref{eq:NJL_energy2}) still 
satisfies the property (\ref{eq:property}).
This can be understood as follows:
the effective potential in the MFA is given by a summation of ring diagrams 
where a number of chiral condensates are attached to one central 
fermion loop. It is only this fermion loop which has the chemical 
potential dependence, $\mu_u$ or $\mu_d$. The contributions of $\mu_u$ 
and $\mu_d$ are decoupled, which leads to the property (\ref{eq:property}).

Therefore, the phase quenching for the free energy and 
chiral condensate, Eqs.~(\ref{eq:property}) and (\ref{eq:equivalence_chiral}), 
is exact in this case as well, as pointed out in Ref.~\cite{Toublan:2003tt}.

\subsection{Linear sigma model}
\label{sec:lsm}

The linear sigma model (L$\sigma$M), also known as the Gell-Mann--Levy model 
\cite{GellMann:1960np}, is another model that describes chiral dynamics of QCD. 
Essentially, this is a bosonized theory of the NJL model with adding
potential terms for meson fields.
The Lagrangian of its $\SU(N_c) \times \U(N_f)_L \times \U(N_f)_R$ 
symmetric generalization is given by
\begin{eqnarray}
\label{eq:Lagrangian_lsm}
{\cal L}_{{\rm L}\sigma{\rm M}}={\cal L}_B+{\cal L}_F, 
\end{eqnarray}
where 
\begin{eqnarray}
{\cal L}_B = \frac{1}{N_c}\left[(\partial_\mu {\pi_A})^2
+(\partial_\mu \sigma_A)^2 \right]
+ U(\sigma_A, \pi_A), 
\end{eqnarray}
\begin{eqnarray}
{\cal L}_F
= \bar{\psi}
\left[\gamma^\mu\partial_\mu + \mu_f \gamma_4
+ \frac{g}{N_c}(\sigma_A + i\gamma^5 {\tau}_A {\pi_A})
\right]\psi,  
\end{eqnarray}
and 
\begin{eqnarray}
U(\sigma_A, \pi_A)
=
\frac{\lambda}{N_c}\left[ {\sigma_A}^2 + {\pi_A}^2 - (N_c f)^2 \right]^2 - H \sigma_0,  
\end{eqnarray}
where $\sigma_A \sim \bar \psi \tau_A \psi$ ($\sigma_0 \sim \bar \psi \psi$) and 
$\pi_A \sim \bar \psi i \gamma_5 \tau_A \psi$ with the
$\U(N_f)$ flavor generators $\tau_A$.
Note that we have included flavor nonsinglet scalars and flavor singlet
pseudoscalar to the conventional L$\sigma$M (see, e.g., Ref.~\cite{Bowman:2008kc})
to maintain the $\U(N_f)_L \times \U(N_f)_R$ symmetry.
The parameters $g,\lambda, f$ and $H$ are taken to be of order $N_c^0$.\footnote{We note 
that our arguments below do not rely on the ansatz for $\lambda$ and $f$, 
as long as they are independent of $N_c$: e.g.,
$\lambda(T)=\lambda_0 \left[1-(T/T_0)^2 \right]$ with some
constants $\lambda_0=O(N_c^0)$ and $T_0=O(N_c^0)$ adopted in Ref.~\cite{Heinz:2011xq}.
Our normalization is related to that in Ref.~\cite{Heinz:2011xq} 
via $\sigma_{\rm ours}=\sqrt{N_c}\sigma_{\rm Heinz}$.}
With this normalization, $\langle \sigma_0 \rangle \sim N_c^1$
in the chiral symmetry broken phase. 
Also the coupling constants involving $\sigma_A$ and $\pi_A$, 
which are physically identified with fermion bilinears 
(mesons), correctly reproduce the usual power counting in the large-$N_c$ QCD.

To argue the orbifold equivalence, we consider the $\SO(2N_c)$ gauge group
counterpart of the L$\sigma$M, whose Lagrangian is given by
\begin{eqnarray}
\label{eq:SO_lsm}
{\cal L}^{\SO}_{{\rm L}\sigma{\rm M}}={\cal L}^{\SO}_B+{\cal L}^{\SO}_F, 
\end{eqnarray}
where 
\begin{eqnarray}
{\cal L}^{\SO}_B &=& \frac{1}{N_c}\left[(\partial_\mu {\pi_A})^2
+(\partial_\mu \sigma_A)^2 + |\partial_\mu d_A^+|^2
+ |\partial_\mu d_A^-|^2 \right] 
\nonumber \\
& & + U(\sigma_A, \pi_A, d_A^+, d_A^-), 
\end{eqnarray}
\begin{eqnarray}
{\cal L}^{\SO}_F
&=& i \bar{\Psi}_F \left(\sigma^\mu \partial_\mu + \mu_F \gamma_4 \right)\Psi_F
+\frac{g}{N_c} \bar \psi (\sigma_A + i\gamma^5 \tau_A \pi_A) \psi
\nonumber \\
& &+ \frac{g}{N_c} \psi^T C (\tau_A \bar d_A^- + i\gamma^5 \tau_A \bar d_A^+) \psi
\nonumber \\
& &+ \frac{g}{N_c} \bar \psi C (\tau_A d_A^- + i\gamma^5 \tau_A d_A^+) \bar \psi^T,  
\end{eqnarray}
and 
\begin{eqnarray}
U(\sigma,\vec{\pi})
=
\frac{\lambda}{N_c}\left[\sigma_A^2 + \pi_A^2 + |d_A^+|^2 + |d_A^-|^2 
- (N_c f)^2\right]^2 - H \sigma_0,
\nonumber \\
\end{eqnarray}
where $d_A^+ \sim \psi^T C i \gamma_5 \tau_A \psi$ and 
$d_A^- \sim \psi^T C \tau_A \psi$ (indices $\pm$ denote the parity)
and $\Psi_F$ is defined in Eq.~(\ref{eq:large_fermion}).

The orbifold projection from $\SO(2N_c)$ L$\sigma$M$_B$ to
$\SU(N_c)$ L$\sigma$M$_B$ can be defined as follows:
the projection for fermions is the same as Eq.~(\ref{eq:fermion_baryon}),
\beq
\psi^{\rm SO}_{a} &=& \omega (J_c)_{aa'} \psi^{\rm SO}_{a'},
\eeq
and the projection for mesons is to throw away $d_A^+$ and $d_A^-$
from $\SO(2N_c)$ L$\sigma$M$_B$.
It is easy to see that this projection maps $\SO(2N_c)$ L$\sigma$M$_B$ 
into $\SU(N_c)$ L$\sigma$M$_B$.

Let us now consider how the orbifold equivalence can be shown 
within this model.
The orbifold equivalence of the fermionic part is simple: because
the fermionic part of the L$\sigma$M can be regarded as fermions 
in the presence of a background field $\sigma_A$, $\pi_A$, $d_A^+$,
and $d_A^-$, the large-$N_c$ equivalence holds as long as 
the background field does not break the projection symmetry
(i.e., outside the diquark condensation phase).

To understand the orbifold equivalence in the bosonic sector,
we consider a neutral meson-meson scattering 
between $\SO(2N_c)$ L$\sigma$M$_B$ and 
$\SU(N_c)$ L$\sigma$M$_B$ as an example.
(A similar argument can be found in Ref.~\cite{Cherman:2011mh}.)
The generalizations to general scattering amplitudes and to 
the case between $\SO(2N_c)$ L$\sigma$M$_I$ and 
$\SU(N_c)$ L$\sigma$M$_B$ are straightforward. 

First note that a meson loop is absent in the large-$N_c$ limit (MFA);
we can concentrate on tree-level scatterings where
the external lines are neutral mesons. The
external legs are neutral pions in $\SO(2N_c)$ L$\sigma$M$_B$
and are $\pi^0$s in $\SU(N_c)$ L$\sigma$M$_B$ (see Tab.~\ref{tab:example}).
Because the neutral-meson coupling constants are taken 
to be the same between the two, a possible difference of 
the scattering amplitude comes from appearance 
of charged mesons (diquarks and antidiquarks)
in the internal lines of the scattering diagram in $\SO(2N_c)$ L$\sigma$M$_B$,
whose counterparts do not exist in $\SU(N_c)$ L$\sigma$M$_B$.
However, this is impossible due to the conservation of the global
${\mathbb Z}_2$ charge, as long as the ${\mathbb Z}_2$ symmetry 
is not broken spontaneously: when the external legs are neutral mesons,
mesons must also be neutral in the internal lines of the diagram.
One might still suspect that a pair of diquarks (antidiquarks), which
is neutral under ${\mathbb Z}_2$, could appear in the diagram. 
But this necessitates a meson loop and is suppressed in the large-$N_c$ limit.
Hence, the neutral meson scattering amplitude must agree between the two.

From the equivalence in both fermionic and bosonic sectors,
the equivalence holds in the full theories between 
$\SO(2N_c)$ and $\SU(N_c)$ L$\sigma$M$_B$ in the MFA, 
as long as the projection symmetry is unbroken.
One can similarly show the equivalence between 
$\SO(2N_c)$ L$\sigma$M$_B$ and $\SU(N_c)$ L$\sigma$M$_I$.
Therefore, the phase quenching is exact in the L$\sigma$M in the MFA.

Using the similar argument given in Sec.~\ref{sec:instanton},
one can also show the phase quenching in the conventional L$\sigma$M
with the $\SU(N_c) \times \SU(N_f)_L \times \SU(N_f)_R \times \U(1)_B$ symmetries. 
The Lagrangian is given by 
\begin{equation}
{\cal L}_{{\rm L}\sigma{\rm M}}={\cal L}_B+{\cal L}_F,
\end{equation}
where
\begin{eqnarray}
{\cal L}_B = \frac{1}{N_c}\left[(\partial_\mu {\pi^a})^2
+ (\partial_\mu\sigma)^2\right]
+ U(\sigma, {\pi^a}), 
\end{eqnarray}
\begin{eqnarray}
{\cal L}_F
= \bar{\psi}
\left[\gamma^\mu\partial_\mu
+ \frac{g}{N_c}(\sigma+i\gamma^5 {\tau^a}{\pi^a})
\right]\psi,  
\end{eqnarray}
and 
\begin{eqnarray}
U(\sigma,{\pi^a})
=\frac{\lambda}{N_c}\left[\sigma^2 + ({\pi^a})^2-(N_c f)^2\right]^2 - H \sigma,
\end{eqnarray}
where $\pi^a \sim \bar \psi i\gamma_5 \tau^a \psi$ with $\tau^a$ being the $\SU(N_f)$ generators.
In the MFA, the effective potential is given by a summation of diagrams that have 
one central fermion loop with a number of meson fields $\sigma$ and $\pi^a$ attached. 
It is again only this fermion loop which depends on the chemical potential; 
the contributions of $\mu_u$ and $\mu_d$ are decoupled, 
and the phase quenching is exact in the MFA.

\subsection{Polyakov--Nambu--Jona-Lasinio model}
The Polyakov-Nambu-Jona-Lasinio (PNJL) model \cite{Fukushima:2003fw, Megias:2004hj, Ratti:2005jh}
is an extended version of the NJL model by adding Polyakov-loop 
degrees of freedom to account for the confinement/deconfinement. 
The Polyakov loop (expectation value) is defined by
\beq
{\ell}=\frac{1}{N_c}\langle \tr L \rangle, \qquad
{\bar \ell}=\frac{1}{N_c}\langle \tr L^{\dag} \rangle.
\eeq
Here $L$ is an $N_c \times N_c$ color matrix
\beq
L({\bf x}) ={\cal P} \exp \left[i \int_0^{\beta} d \tau 
A_4({\bf x}, \tau) \right],
\eeq
with ${\cal P}$ being the path ordering, $A_4=iA_0$, and $\beta=1/T$.
The Lagrangian of the PNJL model is given by \cite{Fukushima:2003fw, Ratti:2005jh}
\beq
\label{eq:Lagrangian_PNJL}
{\cal L}_{\rm PNJL} &=& 
{\cal L}_{\rm kin} + {\cal L}_{\rm int} + {\cal L}_{\rm pot},\\
{\cal L}_{\rm kin} &=& \bar \psi (\gamma^{\mu}D_{\mu} + \mu \gamma^4) \psi,
\eeq
where the interaction term ${\cal L}_{\rm int}$ is taken to be the same
as that of the NJL model, e.g., Eq.~(\ref{eq:4-fermi}).
On the other hand, $\partial_4$ in the kinetic term of the NJL model 
is replaced by the covariant derivative 
$D_4=\partial_4-iA_4$ in ${\cal L}_{\rm kin}$
(other derivatives are untouched, $D_i = \partial_i$)
and the Polyakov loop potential 
${\cal L}_{\rm pot}={\cal L}_{\rm pot}(\ell, \ell^*, T)$ is introduced. 
The parameters of the Polyakov-loop potential are determined 
by fitting lattice simulation data at $\mu=0$ and finite $T$, but the 
detailed form of the potential is irrelevant in this paper.
In order for the Polyakov loop to take the same expectation value between
$\SO(2N_c)$ theory and QCD at any $T$ (at $\mu=0$) as required by 
the large-$N_c$ equivalence in Sec.~{\ref{sec:orbifolding}}, 
we assume the same potential between the two theories.
(This should be so at the level of MFA as well as one-meson-loop corrections, 
see the remark on $1/N_c$ corrections to gluonic operators 
at the end of Sec.~\ref{sec:orbifolding}.)
This is a necessary input for the model to be consistent with 
the underlying gauge theories.
Once this assumption is made, the equivalence in the gauge sector of the 
model is trivially valid. 

On the other hand, the fermionic part of the PNJL model can be regarded as the NJL model 
in the presence of a background field $A_4$. As long as the background field 
does not break the projection symmetry (and it must be so because the Polyakov-loop 
potential is chosen at $\mu=0$ where the projection symmetry is not broken), 
the equivalence in the fermionic sector is also satisfied from the argument in 
Sec.~\ref{sec:njl}.

Actually, as noted in Ref.~\cite{Sakai:2010kx}, 
the effective potential of the PNJL model satisfies the relation
\begin{equation}
\Omega_{\rm PNJL}(\mu_B)|_{\mu_I=0} = \Omega_{\rm PNJL}(\mu_I)|_{\mu_B=0},
\end{equation}
outside the pion condensation phase in the MFA;\footnote{There is an ambiguity
to take the MFA of the PNJL model in the literature: its effective potential is 
complex at $\mu \neq 0$, and thus, $\ell$ and $\bar \ell$ are independent in general.
However, the orbifold equivalence in the underlying QCD predicts that the 
Polyakov loop in QCD$_B$ agree with that in QCD$_I$ outside the pion condensation
phase where $\ell = \bar \ell$. Therefore, the correct MFA in the PNJL model
must satisfy $\ell = \bar \ell$ in QCD$_B$ in that region.
This is consistent with the claim in Ref.~\cite{Rossner:2007ik}.}
the phase quenching is exact for the free energy, chiral condensate, 
and Polyakov loop in this model. 
 
\subsection{Polyakov-quark-meson model}
One can also consider the extended version of the linear sigma model
by taking into account Polyakov loop degrees of freedom.
This is known as the Polyakov-quark-meson (PQM) model \cite{Schaefer:2007pw}.
The Lagrangian is
\beq
{\cal L}_{\rm PQM}={\cal L}_{{\rm L}\sigma{\rm M}} + {\cal L}_{\rm pot},
\eeq
where ${\cal L}_{{\rm L}\sigma{M}}$ is the Lagrangian 
given in Eq.~(\ref{eq:Lagrangian_lsm}) with $\partial_4$ 
replaced by the covariant derivative $D_4=\partial_4-iA_4$
and ${\cal L}_{\rm pot}$
is the same potential of Polyakov loop used in Eq.~(\ref{eq:Lagrangian_PNJL}).
The proof of the L$\sigma$M can be extended straightforwardly, 
just as the proof of the NJL can be extended to that of PNJL.  

\subsection{Chiral random matrix model} 
In this subsection, we explain the orbifold equivalence of 
the chiral random matrix model ($\chi$RMM) following Ref.~\cite{Hanada:2011ju}.
The partition function of the $\chi$RMM \cite{Shuryak:1992pi} 
(for a review, see, e.g., Ref.~\cite{Verbaarschot:2000dy})
is given by an integral over a Gaussian random matrix ensemble,
\beq
\label{eq:RMM}
Z=\int d\Phi \prod_{f=1}^{N_f} \det {\cal D}_f \ 
e^{-\frac{N \beta}{2}G^2 \tr \Phi^{\dagger} \Phi},
\eeq
where $\Phi$ is an $N \times N$ random matrix element. 
The parameter $G$ is a normalization of the Gaussian.
This theory does not have spacetime dependence;
the size of the matrix $N$ corresponds to the spacetime volume 
which is taken infinity in the end (thermodynamic limit).
 
The matrix structure of the Dirac operator ${\cal D}$ is chosen 
such that it reproduces correct anti-unitary symmetries and 
global symmetry breaking pattern of the system.
We can also add the quark mass $m$, quark chemical potential $\mu$ 
\cite{Stephanov:1996ki}, and temperature $T$ \cite{Halasz:1998qr, Vanderheyden:2005ux} 
into ${\cal D}$. At $T=0$, the Dirac operator is written as
\begin{eqnarray} 
{\cal D}_f = \left(
\begin{array}{cc}
m_f \textbf{1}& \Phi+\mu_f\textbf{1} \\
-\Phi^\dagger +\mu_f\textbf{1} & m_f \textbf{1}
\end{array}
\right).
\end{eqnarray}
Here the matrix $\Phi$ is taken to be real, complex or quaternion real. 
Each case is respectively characterized by the Dyson index 
$\beta=1$, $\beta=2$ or $\beta=4$, which represents independent degrees 
of freedom per each matrix element.
$\beta=1$ corresponds to $\SU(2)$ QCD and $\Sp(2N_c)$ gauge theory, 
$\beta=2$ to QCD with $N_c\ge 3$, 
$\beta=4$ to QCD with adjoint fermions and $\SO(2N_c)$ gauge theory.  

Because the $\chi$RMM is a large-$N$ (not large-$N_c$) matrix model, 
one can prove the orbifold equivalence as in the same way as the field theories, 
just by replacing $N_c$ with $N$. 
In the following, we construct the orbifold projection from 
$\beta=4$ RMM at finite $\mu_B$ ($\beta=4$ RMM$_B$) to 
$\beta=2$ RMM at finite $\mu_B$ ($\beta=2$ RMM$_B$) 
or $\beta=2$ RMM at finite $\mu_I$ ($\beta=2$ RMM$_I$) at $T=0$,
which can easily be generalized to nonzero $T$ (for the orbifold projection 
from $\beta=1$ to $\beta=2$, see Ref.~\cite{Hanada:2011ju}.)
The construction of the orbifold projections is almost the same as 
the projections from $\SO(2N_c)_B$ to QCD$_B$ or QCD$_I$.
For simplicity, we consider degenerate quark masses $m_f=m$.  

The partition function of the $\beta=4$ RMM$_B$ 
is given by 
\begin{eqnarray}
Z=\int d\Phi d\Psi \ e^{-S}, \qquad S=S_{B}+S_{F},
\end{eqnarray}
with
\begin{eqnarray}
S_B
=\frac{N\beta}{2}G^2 \tr \Phi^\dagger \Phi,
\end{eqnarray}
and
\begin{eqnarray}
\label{eq:RMM-action}
S_F
=
\sum_{f=1}^{N_f}
\bar{\Psi}_f {\cal D}\Psi_f, \quad
{\cal D}
=
\left(
\begin{array}{cc}
m \textbf{1}_{2N}& \Phi+\mu\textbf{1}_{2N} \\
-\Phi^\dagger +\mu\textbf{1}_{2N} & m \textbf{1}_{2N}
\end{array}
\right),
\nonumber \\
\end{eqnarray}
where $\Phi$ is a $2N\times 2N$ quaternion real matrix and
$\Psi_f$ are complex $4N$-component fermions. 
They can be decomposed as
\begin{eqnarray}
\label{eq:quaternion}
\Phi &\equiv & \sum_{\mu=0}^3 a^{\mu} i \sigma_{\mu}=
\left(
\begin{array}{cc}
a^0 + i a^3 &  a^2 + i a^1 \\
-a^2 + i a^1 & a^0 - i a^3
\end{array}
\right),
\\
\Psi &\equiv &
\left(
\begin{array}{c}
\psi_R \\ \psi_L
\end{array}
\right), \quad 
\psi_{R,L}= \left(
\begin{array}{c}
\xi_{R,L} \\ 
\zeta_{R,L}
\end{array} \right).
\end{eqnarray}
Here $\psi_{R,L}$ are $2N$-component fermions which are 
further decomposed into $N$-component fermions 
$\xi_{R,L}$ and $\zeta_{R,L}$, and 
$a^{\mu}$ are $N\times N$ real matrices.

In order to obtain $\beta=2$ RMM$_B$, 
we impose the projection condition as 
\begin{eqnarray}
J \Phi J^{-1} = \Phi, \qquad
\psi_{R,L} = \omega J \psi_{R,L},
\end{eqnarray}
where $J = -i \sigma_2 \otimes 1_{N}$ and 
$\omega=e^{i\pi/2}$ as defined before. 
Then it is easy to see $\beta=2$ RMM$_B$ is obtained after the projection. 
In the same way, $\beta=2$ RMM$_I$ is obtained by using 
\beq
J \Phi J^{-1} = \Phi, \qquad
J\psi_{R,L} J_i^{-1}=\psi_{R,L}, 
\eeq 
where $J_i$ acts on the flavor indices. 

Let us check the exactness of the phase quenching in this model
which is already observed in Ref.~\cite{Klein:2003fy}.
The effective potential of two-flavor $\beta=2$ RMM with
the quark chemical potentials $\mu_u$ and $\mu_d$ 
is computed, using the saddle point approximation 
for $N \rightarrow \infty$ as \cite{Klein:2003fy}:
\beq
\label{eq:potential2}
\Omega_{\rm RMM} &=& G^2[(\sigma_u-m_u)^2+(\sigma_d-m_d)^2 + 
2(\rho-\lambda)^2]
\nonumber \\
	& & -\frac{1}{2}\sum_{\pm}
\ln [(\sigma_u + \mu_u \pm iT)(\sigma_d - \mu_d \mp iT) + \rho^2]
\nonumber \\
& & \qquad \quad
\times [(\sigma_u - \mu_u \mp iT)(\sigma_d + \mu_d \pm iT) + \rho^2].
\nonumber \\
\eeq
The chiral condensate and pion condensate are related to
$\sigma_{u,d}$ and $\rho$ as
\beq
\langle \bar u u \rangle
&=& \left. \frac{1}{2N}\partial_{m_u}\ln Z \right|_{m_u=0} = - G^2 \sigma_u,
\\
\langle \bar d \gamma^5 u \rangle 
&=& \left. \frac{1}{4N}\partial_{\lambda}\ln Z \right|_{\lambda=0} = - G^2 \rho.
\eeq
Outside the pion condensation phase, $\rho=0$, 
the potential (\ref{eq:potential2}) satisfies the relation:
\beq
\Omega_{\rm RMM}(\mu_B)|_{\mu_I=0}
=\Omega_{\rm RMM}(\mu_I)|_{\mu_B=0}.
\eeq
By differentiating with respect to the quark mass, it follows that
the chiral condensates are identical between RMM$_B$ and RMM$_I$ 
for $\rho=0$.

\subsection{Strong-coupling lattice QCD} 
The orbifold equivalence can be extended to the strong-coupling 
expansion of lattice QCD. For clarity, consider the action 
on the lattice with staggered fermions in the chiral limit \cite{Kogut:1974ag}:
\begin{align}
S[U, \chi, \bar \chi ] &= S_G[U] + S_F [U, \chi, \bar \chi ], 
\nonumber \\
S_G[U] &= \frac{2N_c}{g^2} \sum_{x,\mu,\nu}\left[1-\frac{1}{N_c}{\rm Re}\tr U_{\mu \nu}(x) \right],
\end{align}
\begin{align}
S_F [&U, \chi, \bar \chi] \nonumber \\
&=\frac{1}{2}\sum_x  \eta_0(x)
\left[\bar \chi(x)e^{\mu}U_0(x) \chi(x + \hat 0) \right. \nonumber \\
& \left. - \bar \chi(x + \hat 0)e^{-\mu}U_0^{\dag}(x)\chi(x) \right] 
+ \frac{1}{2}\sum_x \sum_{j=1}^d \eta_j(x) \nonumber \\
& \times \left[\bar \chi(x)U_j(x) \chi(x + \hat j) 
- \bar \chi(x + \hat j)U_j^{\dag}(x)\chi(x) \right].
\end{align}
Here
\beq
U_{\mu \nu}(x)=U_{\nu}^{\dag}(x)U_{\mu}^{\dag}(x+ \hat \nu)U_{\nu}(x + \hat \mu)U_{\mu}(x),
\eeq
is the plaquette, $U_{\mu}$ is the $\SU(N_c)$ gauge link variable,
$\chi$ is the fermion field, $\eta_{\mu}(x)$ is defined as
$\eta_0(x)=1$ and $\eta_j(x)=(-1)^{\sum_{i=1}^j x_{i-1}}$,
and $d$ is the number of spatial directions. 

In the strong-coupling limit $g \rightarrow \infty$, the gluon action $S_G$
can be dropped, and the theory is given just by $S_F$. 
Because the orbifold projection can be defined for the resultant action 
at the lattice level, the equivalence immediately follows in the large-$N_c$ limit. 
Note here that the large-$N_c$ limit is taken for the action $S$ 
{\it after} the strong-coupling limit $g \rightarrow \infty$.
In this case, however, the leading order in $1/N_c$ expansion
does not correspond to the MFA in the literature \cite{Nishida:2003fb}. 

The phase quenching is also exact at large $N_c$ to the next-to-leading order (NLO) 
and higher order in $1/g^2$. We again note that the large-$N_c$
limit is taken {\it after} we truncate into the NLO (or higher) action 
of the strong-coupling lattice QCD. 
(For the higher-order calculation in $1/g^2$, see, e.g., Ref.~\cite{Nakano:2009bf}.)

\section{Equivalences in holographic models of QCD}
\label{sec:holography}
In this section, we apply the orbifold equivalence
to holographic models of QCD.
Since the holography (or the gauge/gravity duality) maps 
a four-dimensional (4D) strongly-coupled gauge theory to 
a five dimensional classical gravity theory
in the large $N_c$ and large 't Hooft coupling limits, we expect that we can use
the large-$N_c$ orbifold equivalence in holographic models,
as originally proposed in Ref.~\cite{Kachru:1998ys}. 
We consider below are the D3/D7 model
and the Sakai-Sugimoto model.
 
\subsection{D3/D7 model} 
In this section we explain the equivalence in the D3/D7 model 
\cite{Karch:2002sh} following Ref.~\cite{Hanada:2012nj}.
Let us start with 4D ${\cal N}=4$ $\U(2N_c)$ supersymmetric Yang-Mills theory, 
which is realized around a stack of $2N_c$ D3-branes. 
The massless spectrum of D3-branes involves a vector multiplet on the worldvolume 
$A_{0123}$ and three complex scalar multiplets describing the transverse motion $X_{45}$, $X_{67}$, $X_{89}$. 
At large $N_c$, this system is described by the type IIB superstring on $AdS_5\times S^5$.  
In order to introduce $2N_f$ fundamental matters, we add $2N_f$ D7-branes, 
which wrap on $S^3\subset S^5$ \cite{Karch:2002sh}: 
$$
\begin{array}{r|cccccccccc}
 \! & 0 & 1 & 2 & 3 & 4 &5 & 6& 7& 8 & 9 \\
\hline {\rm D3} & \circ & \circ & \circ & \circ & \cdot & \cdot & \cdot & \cdot & \cdot & \cdot \\
{\rm D7} & \circ & \circ & \circ & \circ & \cdot & \cdot & \circ & \circ & \circ & \circ
\end{array}
$$
Then two $2N_c\times 2N_f$ chiral multiplets $H^A$ describing strings from D3 to D7-branes and the reversed strings ${\widetilde H}_A=\epsilon_{AB} {H^B}^\dagger$ emerge. 
In the large-$N_c$ limit with fixed $N_f$, one can neglect the back reaction and treat the D7-branes 
as probes in $AdS_5\times S^5$ background.\footnote{Note that the probe approximation in holography 
is different from the quenched approximation in QCD in general; while the latter does not 
distinguish $\mu_B$ and $\mu_I$ as we have seen in Sec.~\ref{sec:quenched_approximation},
the former does distinguish them as we will see below.}
By writing the $AdS_5\times S^5$ metric as
\begin{equation}
ds^2=\frac{|y|^2}{R^2}\eta_{\mu\nu} dx^\mu dx^\nu+\frac{R^2}{|y|^2}\sum_{i=4}^9 dy_i^2,
\end{equation}
the D7 are localized at $y_8=y_9=0$ and extend along all the other directions. 
Then open strings connecting D3 and D7 provide us with $\U(N_c)\times \U(N_f)$ 
bi-fundamental matters, which resemble the $\U(N_c)$ fundamental matters 
with $\U(N_f)$ flavor symmetry.  
The dynamics of quarks and mesons is described by the Dirac-Born-Infeld (DBI) action 
\begin{eqnarray}\label{eq:DBI}
S_{\rm DBI}=-T_7\int d^8\xi\,\ {\rm Tr} \sqrt{-\det\left(G+2\pi\alpha'F\right)},
\end{eqnarray}
where $\xi$ are the world-volume coordinates, 
$G$ is the pull-back of the spacetime metric to the world volume and 
$F$ is the field strength of the gauge fields on the brane, and $T_7$ is the D7-brane tension. 
The chemical potential can be introduced as a background field 
of zeroth component of the gauge field on D7-branes. 
Here we choose the isospin chemical potential,   
\begin{eqnarray}
A_0^{\rm background}
=
i\mu J_{2N_f}, 
\end{eqnarray}
where $J_{2N_f}=-i\sigma_2\textbf{1}_{N_f}$. 
Starting with this theory,  
one can obtain an $\SO(2N_c)$ theory with $N_f$ flavors at finite 
$\mu_B$ via an orientifold projection, 
and a $\U(N_c)$ theory with $N_f$ flavors at finite $\mu_B$
by further orbifold projection.   
For the orientifold projection, we introduce an O7-plane and a ${\mathbb Z}_2$ singularity as follows: 
$$
\begin{array}{r|cccccccccc}
 \! & 0 & 1 & 2 & 3 & 4 &5 & 6& 7& 8 & 9 \\
\hline {\rm D3} & \circ & \circ & \circ & \circ & \cdot & \cdot & \cdot & \cdot & \cdot & \cdot \\
{\rm O7/D7} & \circ & \circ & \circ & \circ & \cdot & \cdot & \circ & \circ & \circ & \circ \\
\mathbb{Z}_2 & \circ & \circ & \circ & \circ & \circ & \circ & \cdot & \cdot & \cdot & \cdot
\end{array}
$$
The geometric effect of the ${\mathbb Z}_2$ action is a reflection in the transverse directions $x^{6,7,8,9}$.
Hence the orientifold projection for the fields on D3 is 
\begin{equation}
\begin{array}{rcl}
A'_{0123} & = & -\; (A'_{0123})^T\; , \\[3pt]
X'_{45} & = & -\; (X'_{45})^T\;  , \\[3pt]
X'_{67,89} & = & \; (X'_{67,89})^T\; .
\end{array}
\end{equation}
Therefore, the orientifold projection for the gauge field is
\begin{equation}\label{eq:projomega}
A'_\mu = \tfrac{1}{2}\left(A_\mu-A_\mu^T \right),
\end{equation}
so the projected gauge field is antisymmetric and spans an $\SO(2 N_c)$ algebra. The field $X_{45}$ is in an antisymmetric (adjoint) representation, while for the fields $X_{67,89}$ the orientifold action projects them to a symmetric representation. 
Open strings connecting D3 and D7 are projected as 
\begin{eqnarray}
H^{\prime A} & = & - i \epsilon_{AB} \left( H^{\prime B} J_{2N_f}^{-1}\right)^*,  
\end{eqnarray}
and the fields on D7 are projected as 
\begin{eqnarray}
A'_{0123} & = &  - J_{2N_f}\; (A'_{0123})^T\; J_{2N_f}^{-1}, \\[3pt]
X'_{45} & = & - J_{2N_f}\; (X'_{45})^T\; J_{2N_f}^{-1},  \\[3pt]
A'_{6789} & = &  -J_{2N_f}\; (A'_{6789})^T\; J_{2N_f}^{-1}.
\end{eqnarray}
The chemical potential remains unchanged, 
\begin{eqnarray}
(A'_0)^{\rm background}
=
i\mu J_{2N_f}.  
\end{eqnarray}
It can be regarded as both $\mu_B$ and $\mu_I$, 
because there is no difference between the two in the $\SO(2N_c)$ theory. 

By further performing a ${\mathbb Z}_2$ orbifold projection, 
one obtains a $\U(N_c)$ theory with $N_f$ flavors at finite $\mu_B$. 
The $\mathbb{Z}_2$ projection is 
\begin{equation}
\begin{array}{rcl}
 A''_{0123} & = &  J_{2N_c}\; A''_{0123}\; J_{2N_c}^{-1}, \\[3pt]
 X''_{45} & = &  J_{2N_c}\; X''_{45}\; J_{2N_c}^{-1}, \\[3pt]
  X''_{67,89} & = & -J_{2N_c}\; X''_{67,89}\; J_{2N_c}^{-1}, 
\end{array}
\end{equation}
for D3-D3 strings,  
\begin{eqnarray}
H^{\prime\prime A} &=& J_{2N_c} H^{\prime\prime A} J_{2N_f}^{-1},
\end{eqnarray}
for D3-D7 strings and 
\begin{eqnarray}
A''_{0123}& = & J_{2N_f}\; A''_{0123}\; J_{2N_f}^{-1},\\[3pt]
X''_{45} & = & J_{2N_f}\; X''_{45}\; J_{2N_f}^{-1}, \\[3pt]
A''_{6789} & = & -J_{2N_f}\; A''_{6789}\; J_{2N_f}^{-1},
\end{eqnarray} 
for D7-D7 strings.  
The background field turns into the one corresponding to $\mu_B$. 
The dual gravity geometry changes to $AdS_5\times {\mathbb R}P^5$ through these projections. 
D7's, which were wrapping on $S^3\subset S^5$ before the projections, wrap on ${\mathbb R}P^3\subset{\mathbb R}P^5$. 
Because $S^3$ and ${\mathbb R}P^3$ are locally the same, the DBI actions are the same 
except that the integration is performed on ${\mathbb R}P^3$ instead of $S^3$ and 
the gauge fields are restricted to satisfy the projection condition.\footnote{As is 
well-known, the DBI action has an ambiguity of the ordering of matrix variables. 
Here we assume there is a right ordering (though we do not know it explicitly) 
and the projections do not affect that ordering.} 
Therefore the equations of motion derived from the DBI actions are the same unless the solutions in 
$\U(2N_c)$ and $\SO(2N_c)$ theories break 
the projection symmetry (or equivalently, if the solution is invariant under the projection symmetry), 
and hence the large-$N_c$ equivalence holds. 

\begin{figure}
\begin{center}
\includegraphics[width=8cm]{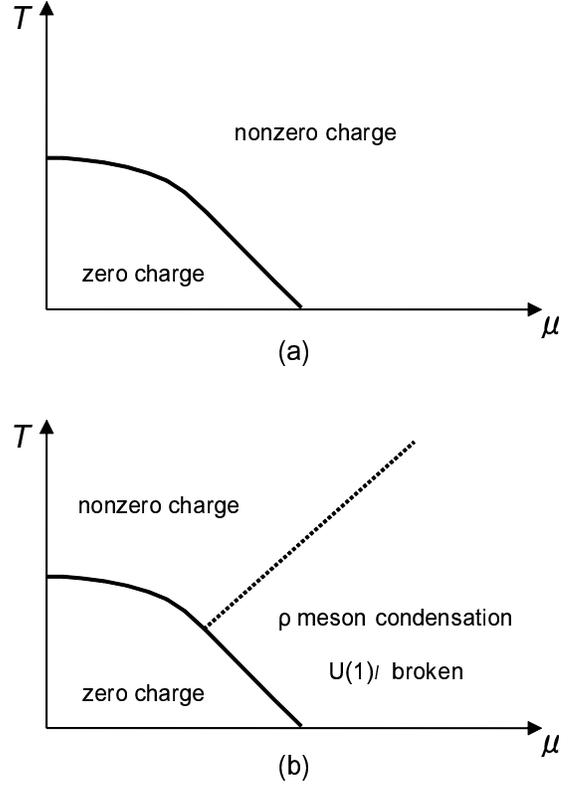}
\end{center}
\vspace{-0.5cm}
\caption{Phase diagram of the D3/D7 model (a) with $\mu_B$ \cite{Mateos:2007vc} 
and (b) with $\mu_I$ \cite{Erdmenger:2008yj,Ammon:2008fc,Erdmenger:2011hp}
with a nonzero quark mass.  
The equivalence holds outside the rho meson condensation phase.}
\label{fig:D3D7}
\end{figure}

The phase diagrams of the D3/D7 models with $\mu_B$ \cite{Mateos:2007vc}
and $\mu_I$\cite{Erdmenger:2008yj,Ammon:2008fc,Erdmenger:2011hp} 
have been studied.\footnote{Notice that the $\SU(N_c)$ theory with $\mu_B$ we are considering, 
which is dual to the string theory on $AdS_5\times {\mathbb R}P^5$, 
is slightly different from the one studied in Ref.~\cite{Mateos:2007vc}, 
which contains three complex adjoint scalars and is dual to the string theory on $AdS_5\times S^5$. 
However these theories are equivalent in the large-$N_c$ limit 
and the solution to the classical equations of motion are the same.}
Schematic picture of the phase diagrams with a nonzero quark mass is shown in Fig.~\ref{fig:D3D7}. 
There are phases with no charge density (analogous to gluon plasma and a gas of mesons) and 
with nonzero charge density (quark-gluon plasma). With $\mu_I$, there is yet another phase,
a rho meson condensation where $\U(1)_I$ symmetry is broken 
(analogous to the pion condensation phase in QCD$_I$). 
In that region the equivalence does not hold. 

\subsection{Sakai-Sugimoto model} 
Sakai-Sugimoto model \cite{hep-th/0412141} has reproduced the low-energy hadron spectrum successfully. 
It has also been used to study the chiral phase transition 
(see, e.g., Refs.~\cite{Aharony:2006da, Horigome:2006xu, Parnachev:2007bc}).  
It consists of $N_c$ D4-branes wrapping on a compactified circle, 
$N_f$ D8-branes and $N_f$ anti-D8-branes. 
Gauge symmetries on D8 and anti-D8 are identified as flavor symmetries 
$\U(N_f)_L$ and $\U(N_f)_R$, respectively. 
When $N_f/N_c\ll 1$,  D8 and anti-D8 can be treated as probes on the D4 background. 
In this setup it has been shown that a D8-brane and an anti-D8-brane merge to form single D8-brane, 
so that $N_f$ D8-branes remain and $\U(N_f)_L\times \U(N_f)_R$ 
is broken down to $\U(N_f)_V$. This is the geometric realization of 
the spontaneous chiral symmetry breakdown.  
The flavor dynamics such as the meson spectrum can be read off from the DBI action for D8-branes 
and the Chern-Simons term.  

As in the D3/D7 model, we consider the probe (anti-)D8-branes 
in the D4-brane background. 
In the chiral symmetry breaking phase, 
the D4-brane background is given by 
\begin{align}
 ds^2 =& 
  \left(\frac{U}{R}\right)^{3/2} 
  \left(
   \eta_{\mu\nu} dx^\mu dx^\nu + f(U) (dx^4)^2
  \right)
\nonumber \\  
&+\left(\frac{R}{U}\right)^{3/2}
  \left(
   \frac{dU^2}{f(U)} + U^2 d\Omega_4^2
  \right) , 
\end{align}
where $f(U) = 1- U_{KK}^3/U^3$. 
Here $U_{KK}$ is a constant which is related to the radius of the compactified dimension. 
D4-branes are extended along $x^\mu$ and $x^4$ directions, where 
$x^\mu$ with $\mu = 0,1,2,3$ is 4 dimensional spacetime and 
$x^4$ direction is compactified to $S^1$. 
D8-branes are embedded in $x^\mu$, $z$, and $S^4$ directions, 
where $z$-direction is one dimensional space embedded in $(U,x^4)$ space. 
Extra-dimensions of $S^4$ are integrated out, 
and gauge fields on D8-branes $A_\mu$ and $A_z$ are 
related to (axial-)vector mesons and a pseudoscalar pion, respectively. 
Here, we explain the orbifold equivalence in this setup. 
It is straightforward to extend the equivalence to 
other applications of the Sakai-Sugimoto model, 
for example, analysis of chiral phase structure, 
as long as it does not depend on the details of extra-dimensions. 

The equivalence can be shown in a similar fashion 
to the D3/D7 model. 
Normalizable modes of the gauge fields on the D8-branes  
correspond to the quark current and 
non-normalizable modes give their source. 
The chemical potential can be introduced 
as a non-vanishing background for the time component of the gauge fields. 
The background is taken to be proportional 
to $1_{N_f}$ for $\mu_B$, 
and proportional to $\sigma_2 \otimes 1_{N_f/2}$ 
for $\mu_I$
\cite{Horigome:2006xu, Parnachev:2007bc}. 
We start with the Sakai-Sugimoto model with $\mu_I$. 
By performing the orientifold projection 
we obtain ${\rm O}(2N_c)$ analogue of the Sakai-Sugimoto model 
with $\mu_B$. 
By further imposing the orbifold projection on the ${\rm O}(2N_c)$ model, 
we obtain $\SU(N_c)$ model with $\mu_B$. 
Since $\mu_B$ is not compatible with the symmetry for the orientifold projection, 
we cannot start with the Sakai-Sugimoto model with $\mu_B$. 

The orbifold projection $g$, 
\begin{align}
 x^\mu = x^\mu , \qquad x^i = -x^i
\end{align}
acts on the D8 gauge fields as 
\begin{align}
 A_\mu(x^\mu,x^i) &= \gamma(g) A_\mu(x^\mu,-x^i) \gamma^{-1}(g) , 
 \nonumber \\
 A_i(x^\mu,x^i) &= -\gamma(g) A_i(x^\mu,-x^i) \gamma^{-1}(g) , 
\end{align}
and similarly for the scalars. 
By using the gauge symmetry, 
$\gamma(g)$ can be taken as 
$\gamma(g) = \sigma_2\otimes 1_{N_f}$. 
For the orientifold projection, 
the worldsheet reflection is 
taken in addition to the spacetime reflection. 
It takes the transpose of the Chan-Paton factors 
and gives an additional sign for the gauge fields 
(but no additional sign for scalars). 
The orientifold projection $\Omega$ acts as 
\begin{align}
 A_\mu(x^\mu,x^i) &= -\gamma(\Omega) A_\mu^T(x^\mu,-x^i) \gamma^{-1}(\Omega) , 
 \nonumber \\
 A_i(x^\mu,x^i) &= \gamma(\Omega) A_i^T(x^\mu,-x^i) \gamma^{-1}(\Omega) . 
\end{align}
There are two options for the orientifold, 
$\gamma_+(\Omega) = 1_{2N_f}$ and $\gamma_-(\Omega) = J_{2N_f}$ 
up to the gauge transformation, 
which give ${\rm O}(2N_c)$ theory with ${\rm O}(2N_f)$ flavor symmetry  
and $\Sp(2N_c)$ theory with ${\rm Sp}(2N_f)$ flavor symmetry, respectively.

The orientifold projection of Sakai-Sugimoto model is 
studied in Ref.~\cite{arXiv:0907.2968}. 
In this case, we construct ${\rm O}(2N_c)$ model. 
It can be obtained by introducing O6$^+$ planes as 
\begin{center}
 \begin{tabular}{c|cccccccccc}
  &0&1&2&3&4&5&6&7&8&9 \\\hline
  D4&$\circ$&$\circ$&$\circ$&$\circ$&$\circ$&$\cdot$ &$\cdot$ &$\cdot$ &$\cdot$ &$\cdot$  \\
  D8-$\overline{\rm D8}$ &$\circ$&$\circ$&$\circ$&$\circ$&$\cdot$ &$\circ$&$\circ$&$\circ$&$\circ$&$\circ$ \\
  O6$^+$-$\overline{\rm O6}^+$ &$\circ$&$\circ$&$\circ$&$\circ$&$\cdot$ &$\circ$&$\circ$&$\circ$&$\cdot$ &$\cdot$  \\
 \end{tabular}
\end{center}
Then the orientifold projection acts on 
the gauge fields on the D8-branes as 
\begin{align}
 A_\mu(x^\mu,z) &= -\gamma_+(\Omega) A_\mu^T(x^\mu,-z) \gamma_+^{-1}(\Omega) , 
\nonumber \\ 
 A_z(x^\mu,z) &= \gamma_+(\Omega) A_z^T(x^\mu,-z) \gamma_+^{-1}(\Omega) . 
\end{align}

Next we impose the orbifold projection. 
The fixed plane lies in the following directions:
\begin{center}
 \begin{tabular}{c|cccccccccc}
  &0&1&2&3&4&5&6&7&8&9 \\\hline
  D4&$\circ$&$\circ$&$\circ$&$\circ$&$\circ$&&&&& \\
  D8-$\overline{\rm D8}$ &$\circ$&$\circ$&$\circ$&$\circ$&&$\circ$&$\circ$&$\circ$&$\circ$&$\circ$ \\
  O6$^+$-$\overline{\rm O6}^+$ &$\circ$&$\circ$&$\circ$&$\circ$&&$\circ$&$\circ$&$\circ$&& \\
  $\mathbb Z_2$ &$\circ$&$\circ$&$\circ$&$\circ$&$\circ$&&&&$\circ$&$\circ$ \\
 \end{tabular}
\end{center}
The gauge fields on the D4-brane after the compactification 
become those for $\U(N_c)$ symmetry. 
The orbifold acts on the gauge fields on the D8-brane as 
\begin{align}
 A_\mu(x^\mu,x^i) &= \gamma(\mathbb Z_2) A_\mu(x^\mu,-x^i) \gamma^{-1}(\mathbb Z_2) , 
 \nonumber \\
 A_z(x^\mu,x^i) &= \gamma(\mathbb Z_2) A_z(x^\mu,-x^i) \gamma^{-1}(\mathbb Z_2) , 
\end{align}
where $i=5,6,7$. 

We focus on the constant modes on $S^4$. 
By taking the orbifold projection first, 
$\U(2N_f)$ gauge symmetry on the D8-brane is broken to 
$\U(N_f)\times \U(N_f)$, and we obtain two gauge fields $A^1$ and $A^2$ 
for each $\U(N_f)$. 
The orientifold projection imposes the relation for 
these gauge fields as $A_\mu^1(x^\mu,z) = - A_\mu^2(x^\mu,-z)$ and 
$A_z^1(x^\mu,z) = A_z^2(x^\mu,-z)$. 
By imposing the both projections, 
we obtain the same effective theory with half flavors, 
but $\mu_I$ becomes $\mu_B$. 
Therefore, the equivalence holds as long as 
we consider only $\mathbb Z_2$ invariant sectors. 

\section{Numerical evidence of the phase quenching}
\label{sec:lattice}
In this section we look at previous numerical simulations 
which compared QCD$_B$ with QCD$_I$. We shall confirm that 
QCD at large $N_c$ and model calculations in the MFA 
provide us with a good approximation 
for the phase quenching in three-color QCD. 

\subsection{Reweighting}
In Ref.~\cite{deForcrand:2007uz}, QCD$_B$ and QCD$_I$ are studied by 
using the canonical formalism as a function of the number of up quarks, $Q$. 
The result of the former is obtained by the reweighting method \cite{Ferrenberg:1988yz, Barbour:1997ej}.
They use two staggered fermions
(corresponding to degenerate four up and four down quark species)
with the bare quark mass $am=0.14$ on a $8^3 \times 4$ lattice.
The canonical partition function $Z_{\rm C}(Q)$ is obtained from
the grand canonical partition function $Z_{\rm GC}(\mu_I)$ 
via the fugacity expansion,
\beq
Z_{\rm GC}(V, T, \mu_I)= \sum_Q Z_{\rm C}(V,T,Q)e^{Q\mu_I/T},
\eeq
where 
\beq
Z_{\rm GC}(V, T, \mu_I) &=& \int dA  \det D(\mu_I) e^{-S_{\rm YM}},
\\
Z_{\rm C}(V, T, Q) &=& \int dA \ \hat \det_Q e^{-S_{\rm YM}},
\eeq
with $\hat \det_Q$ being the projected determinant 
for the fixed quark number $Q$. From the above relations, 
the quantities $\hat \det_Q$ and $\det D(\mu_I)$ are assumed to be related through
\beq
\det D(\mu_I)=\sum_Q \hat \det_Q e^{Q\mu_I/T}.
\eeq
This relation allows us to extract $\hat \det_Q$, and hence, $Z_{\rm C}(Q)$.
The canonical free energy is then given by $F_C(Q)=-(1/T)\ln Z_C(Q)$.
The canonical partition function and free energy at finite $\mu_B$ can be obtained
in a similar way.

In the right panel of Fig.~1 and the left panel of Fig.~4 of Ref.~\cite{deForcrand:2007uz}, 
the free energy at various temperatures between $0.5 T_c$ and $1.1 T_c$ are plotted 
as functions of $Q$. 
By putting these plots on top of each other, 
one can see a very nice agreement near the critical temperature 
and $Q \lesssim 100$. It clearly shows the validity of the phase quenching. 
It should also be remarked that the corrections are still tiny for $N_f=8$, 
a larger number of flavors than $N_f=2+1$ in the real world 
[remember that the corrections are $O(N_f/N_c)$ from our large-$N_c$ argument 
in Sec.~\ref{sec:orbifolding}]. 

In Ref.~\cite{Sasai:2003py}, three-color and two-flavor QCD$_B$ and QCD$_I$ 
are studied using staggered fermions with the bare quark mass $am=0.05$
on a $8^3 \times 4$ lattice. 
The former is obtained by the phase reweighting from the latter.
The chiral condensate and the Polyakov loop are computed for $a\mu=0.1$ and $a\mu=0.2$, 
and the results of QCD$_B$ and QCD$_I$ agree within numerical errors,
even for the average phase factor $\sim 0.7$.  

\subsection{Imaginary chemical potential method}
The sign problem is absent when the chemical potential 
is pure imaginary, $\mu=i\mu_{\rm img}$ ($\mu_{\rm img}\in{\mathbb R}$)
\cite{Alford:1998sd, de Forcrand:2002ci}.
This fact can be easily realized by an argument similar to the one around 
Eq.~(\ref{eq:positivity}); since the operator $\gamma^{\mu}D^{\mu} + i \mu_{\rm img} \gamma^4$ 
is anti-Hermitian, 
its eigenvalues $\pm i \lambda_n$ are pure imaginary, $\lambda_n \in {\mathbb R}$,
and the measure is positive semi-definite.
Although the imaginary chemical potential is not physical, 
it is useful if observables are analytic in $\mu^2$ around $\mu^2=0$, 
because the values at $\mu^2>0$ (real chemical potential), 
which are difficult to study due to the sign problem, may be obtained 
through an analytic continuation from $\mu^2<0$ (imaginary chemical potential). 
Note however that the analyticity, which is necessary for the analytic continuation, 
can be lost at any phase transition, such as the chiral and deconfinement transitions. 

Our derivation for the large-$N_c$ equivalence in Sec.~\ref{sec:large-Nc} can also 
be applied for the imaginary baryon and isospin chemical potentials, 
$(\mu_u,\mu_d)=(i\mu_{\rm img},i\mu_{\rm img})$ and 
$(\mu_u,\mu_d)=(i\mu_{\rm img},-i\mu_{\rm img})$, without any modification. 
As a result, the chiral condensates $\langle\bar{\psi}\psi\rangle_B$ and $\langle\bar{\psi}\psi\rangle_I$ 
take the same value at finite imaginary potentials as long as the projection symmetries are unbroken.

In Ref.~\cite{Cea:2012ev}, pseudo-critical temperatures of the chiral transition, 
$T_c(\mu)$, in two degenerate staggered fermions and three-color QCD at $\mu^2>0$ 
were exploited by the extrapolations from $\mu^2<0$ 
(for the bare mass $am=0.05$ on a $16^3 \times 4$ lattice). With an ansatz,
\beq
\frac{T_c(\mu)}{T_c(0)}=1 + a_1 \left( \frac{\mu}{\pi T} \right)^2,
\eeq
they found \cite{Cea:2012ev}
\beq
a_1 &=& -0.465(9) \qquad {\rm for } \ \mu_I,
\nonumber \\
a_1 &=& -0.515(11) \qquad {\rm for }\ \mu_B, 
\eeq 
which provide a nice quantitative agreement already at $N_c=3$.
As found from our arguments above, this difference originates from the $1/N_c$ corrections. 

\subsubsection*{Roberge-Weiss periodicity}
At a finite imaginary baryon chemical potential, the grand canonical partition function 
has the Roberge-Weiss (RW) periodicity \cite{PRINT-86-0360 (BRITISH-COLUMBIA)} 
\begin{eqnarray}
Z\left(\frac{\mu_{\rm img}}{T}\right)
=
Z\left(\frac{\mu_{\rm img}}{T}+\frac{2\pi n}{N_c}\right)
\qquad(n\in{\mathbb Z}),     
\end{eqnarray}
which can be understood as a generalization of the center symmetry of the pure Yang-Mills theory; 
actually the Polyakov loop is transformed as $\ell \to e^{2\pi in/N_c} \ell$.    

In the confinement phase ($\ell =0$) 
the ground state also satisfies the RW periodicity.
Therefore, in the large-$N_c$ limit, there is no $\mu_{\rm img}$-dependence, 
and thus, there is no $\mu$-dependence at $\mu>0$ until the phase transition happens. 
This is consistent with an important property of the large-$N_c$ QCD that 
observables are $T$-independent in the confinement phase.\footnote{This 
can be understood as a generalization of the Eguchi-Kawai reduction
\cite{UT-378-TOKYO} stating that observables are independent of the 
size of the compactified direction in the confinement phase. 
If we use it for the compactified $T$ direction, $T$-independence
of observables immediately follows. 
It leads to the $\mu$-independence at finite $T$ for $\mu<\mu_B/N_c$,
because there is no $\mu$-dependence at $T=0$ for $\mu<\mu_B/N_c$.} 
This can also be understood physically from the fact that
$O(N_c^1)$ observables, such as the chiral condensate, 
cannot be affected by thermal excitations of noninteracting mesons and 
glueballs, which have only $O(N_c^0)$ degrees of freedom \cite{Neri:1983ic}. 

In the deconfinement phase ($\ell \neq 0$) the vacuum does not respect 
the RW periodicity and nontrivial $\mu$-dependence can appear. 

\subsection{Taylor expansion method} 
Another common approach to circumvent the sign problem is the Taylor expansion method;  
one expands the expectation value of an observable in power series of $\mu/T$
\cite{Allton:2002zi, Allton:2005gk, Gavai:2008zr}, 
\begin{eqnarray}
\langle{\cal O}\rangle_B=\sum_{n=0}^\infty c_n^B\left(\frac{\mu}{T}\right)^n
\end{eqnarray}
in QCD$_B$ and 
\begin{eqnarray}
\langle{\cal O}\rangle_I=\sum_{n=0}^\infty c_n^I\left(\frac{\mu}{T}\right)^n
\end{eqnarray}
in QCD$_I$. 
Taylor coefficients $c_n^B$ and $c_n^I$, which are functions of the temperature $T$,  
can be determined by the simulation at $\mu=0$. The large-$N_c$ equivalence tells 
that the coefficients agree in the large-$N_c$ limit. 

In Ref.~\cite{Allton:2005gk}, the coefficients $c_2^B$ and $c_2^I$ for the chiral condensate 
and the pressure of the quark-gluon plasma have been calculated\footnote{ For odd $n$, $c_{n}^B$ 
and $c_{n}^I$ vanish, and the first nontrivial $\mu$-dependences appear in 
$c_2^B$ and $c_2^I$. Although $c_{n}^B$ ($n \geq 4$) have been calculated, 
$c_{n}^I$ ($n\ge 4$) have not been calculated in Ref.~\cite{Allton:2005gk}. 
(Note that, for $n\ge 4$, they use the same symbol $c_n^I$ for another quantity.)} 
in three-color and two-flavor QCD.
Their calculations are performed using staggered fermions with the bare quark mass
$am=0.1$ on a $16^3 \times 4$ lattice. 
The coefficients for the pressure are \cite{Allton:2005gk}
\begin{center}
  \begin{tabular}{|c|c|c|}
  \hline 
$T/T_c$ &  $c_2^B$  &  $c_2^I$ \\ \hline 
0.81 & 0.0450(20) & 0.0874(8) \\
0.90 & 0.1015(24) & 0.1551(14) \\
1.00 & 0.3501(32) & 0.3822(26) \\
1.07 & 0.5824(23) & 0.5972(21) \\
1.16 & 0.7091(15) & 0.7156(14) \\
1.36 & 0.7880(11) & 0.7906(9) \\
1.65 & 0.8157(8) & 0.8169(7) \\
1.98 & 0.8230(7) & 0.8250(6) \\
  \hline
  \end{tabular}
\end{center}
Although the difference between $c_2^B$ and $c_2^I$ are not very small
for $T<T_c$ in the chiral symmetry broken (and confined) phase, 
they agree exceptionally well for $T>T_c$.  
This tendency can naturally be understood, as we have argued
in the end of Sec.~\ref{sec:orbifolding}. 
The coefficients for the chiral condensate are shown in the second panel of Fig.~3.6 
of Ref.~\cite{Allton:2005gk}. There the agreement is even better;
the coefficients agree within errors for $T/T_c\ge 0.87$.

\section{Discussions and outlook}
\label{sec:discussion}
In this paper, we have systematically developed the string-inspired 
technique of the orbifold equivalence in effective models, 
holographic models, and lattice methods for QCD.
As a consequence, we provided the criteria for the validity
of the phase quenching, as summarized in Sec.~\ref{sec:introduction}.
The phase quenching does not produce any quantitative difference of 
the chiral and deconfinement phase transitions in the mean-field approximation 
(MFA) and to the one-meson-loop corrections, respectively, 
outside the pion condensation phase of the phase-quenched theory. 

In the pion condensation phase, the orbifold equivalence breaks down.
Also the $1/N_c$ expansion itself may no longer capture the physics 
of real QCD.\footnote{In the large-$N_c$ limit, 
there is no nuclear liquid-gas transition \cite{Torrieri:2010gz}
and no color superconductivity \cite{Deryagin:1992rw} which are expected 
to be realized in real QCD. This necessitates a phase transition as a function of 
$N_c$ from homogeneous to inhomogeneous matter 
(from a nuclear gas or nuclear liquid to a nuclear crystal
\cite{Klebanov:1985qi, Adhikari:2011zf}
and from a color superconductor to a chiral density wave 
\cite{Deryagin:1992rw} or a chiral quarkyonic spiral \cite{chiralspirals}).}
Since the validity of the MFA in model calculations may be tightly connected to the validity 
of the $1/N_c$ expansion as we have seen in Sec.~\ref{sec:MFA}, 
it is possible that the $1/N_c$ expansion as well as the MFA
are not useful inside the pion condensation phase of QCD$_I$.
If so, previous model calculations supporting the existence of the 
QCD critical point (see Ref.~\cite{Stephanov:2007fk} for a review) 
may not be reliable, 
because it was observed only inside the pion condensation phase 
in model calculations under the MFA \cite{Han:2008xj}.
Actually, by utilizing the large-$N_c$ equivalence, it has recently been shown 
that QCD critical point cannot exist outside the pion condensation phase 
in the large-$N_c$ QCD and effective models in the MFA \cite{Hidaka:2011jj}.
Therefore, the effects beyond the leading order in $1/N_c$ or 
those beyond the MFA should be taken into account to describe the realistic dense matter.

Of course it is important to understand the fate of 
the chiral phase transition outside the pion condensation phase.   
For example, one could study the curvature of the chiral critical 
surface \cite{deForcrand:2006pv} away from $\mu \sim 0$, from which we may 
hopefully infer the behavior of the chiral phase transition and 
the (non)existence of the possible QCD critical point at larger $\mu$.
Based on the phase quenching approximation and using the rooted staggered fermions, 
it was numerically suggested in Ref.~\cite{Kogut:2007mz} that the QCD critical point 
does not exist in three-flavor QCD$_B$ for $\mu \simle m_{\pi}/2$.
A more decisive conclusion should be drawn by detailed 
numerical calculations in the future.

\emph{Note added.}---After this work was completed, we were informed
about Ref.~\cite{Armoni:2012jw} which claims to prove the nonperturbative
equivalence between QCD$_B$ and QCD$_I$ at large $N_c$.
However, their argument and their conclusion about the $1/N_c$ corrections
are nothing more than the previous perturbative arguments \cite{Cohen:2004mw, Toublan:2005rq}.
They consider the expansion of the fermion determinant in $1/N_c$ 
around the Yang-Mills vacuum without quarks
(i.e., quenched approximation of QCD$_B$), which is nonperturbatively 
different from QCD$_B$ itself, as argued in 
Sec.~\ref{sec:quenched_approximation} of our present paper.
Note that nonperturbative fermionic condensates do appear in the large-$N_c$ 
limit, as opposed to their claim, which falsifies their assumption.
Note also that the existence of condensates itself does not spoil the equivalence. 
Rather the symmetry realization is the key; for example 
the equivalence holds with a nonzero chiral condensate. 
They also missed quantities for which the phase quenching 
becomes exact at large $N_c$. 

In order to give a nonperturbative proof of the equivalence, 
one has to take into account the symmetry realization, 
as emphasized in Ref.~\cite{Kovtun:2004bz} 
(see also Refs.~\cite{Cherman:2010jj, Hanada:2011ju, Hanada:2012nj}). 
Probably the most promising approach is to use the orbifold equivalence 
\cite{Cherman:2010jj,Hanada:2011ju} and to consider the Hamiltonian dynamics 
of the theory truncated at one-fermion-loop level, 
by closely following Ref.~\cite{Kovtun:2004bz}. 
We defer it to a future problem.

\section*{Acknowledgment} 
The authors would like to thank S.~Aoki, M.~Buchoff, P. de Forcrand,
S.~Hashimoto, Y.~Hidaka, H.~Iida D.~B.~Kaplan, A.~Li, K.~Nagata, 
A.~Nakamura, A.~Ohnishi, and Y.~Sakai for stimulating discussions and comments.
We also thank F. Karsch for his valuable critical comment, 
which made us reconsider the issue of the phase quenching,
and A.~Armoni and A.~Patella for informing us about their paper \cite{Armoni:2012jw} 
and explaining their logic in detail.
M.~H. was supported in part by the National Science Foundation under Grant No. PHY11-25915.
Y.~M. is supported by JSPS Research Fellowships for Young Scientists.
N.~Y. was supported in part by JSPS Postdoctoral Fellowships for Research Abroad
and JSPS Research Fellowships for Young Scientists.  
M.~H. would like to thank the Kavli Institute for Theoretical Physics and the workshop 
``Novel Numerical Methods for Strongly Coupled Quantum Field Theory and Quantum Gravity" 
for hospitality and a financial support.

\end{document}